\documentclass[fleqn,usenatbib]{mnras}

\usepackage{newtxtext,newtxmath}
\usepackage[T1]{fontenc}

\DeclareRobustCommand{\VAN}[3]{#2}
\let\VANthebibliography\thebibliography
\def\thebibliography{\DeclareRobustCommand{\VAN}[3]{##3}\VANthebibliography}

\usepackage{graphicx}	% Including figure files
\usepackage{amsmath}	% Advanced maths commands

\usepackage{amssymb}	% Extra maths symbols

\graphicspath{/figures} % Include graphics from file

\title{Isotopic ratios for C, N, Si, Al, and Ti in C-rich presolar grains from massive stars}

\author[J. Schofield et al.]{
Jordan Schofield$^{1,9}$*, Marco Pignatari$^{2,3,4,6,7}$, Richard J. Stancliffe$^{8}$, Peter Hoppe$^5$
\\
$^1$Department of Physics and Mathematics, University of Hull, HU6 7RX, UK\\
$^{2}$ Konkoly Observatory, Research Centre for Astronomy and Earth Sciences (CSFK), E\"otv\"os Lor\'and Research Network (ELKH),\\ 
Konkoly Thege Mikl\'{o}s \'{u}t 15-17, H-1121 Budapest, Hungary\\
$^{3}$ CSFK, MTA Centre of Excellence, Budapest, Konkoly Thege Mikl\'{o}s \'{u}t 15-17, H-1121, Hungary\\
$^4$E.A.Milne Centre for Astrophysics, University of Hull, HU6 7RX, UK\\
$^5$Max Planck Institute for Chemistry, Hahn-Meitner-Weg 1, 55128 Mainz, Germany\\
$^6$NuGrid Collaboration, \url{http://nugridstars.org}\\
$^7$Joint Institute for Nuclear Astrophysics - Center for the Evolution of the Elements\\
$^8$H. H. Wills Physics Laboratory, University of Bristol, Tyndall Avenue, Bristol BS8 1TL, UK\\
$^9$NHS North of England Commissioning Support Unit, Appleton House, Lanchester Road, Durham, DH1 5XZ, UK
}

\date{Accepted XXX. Received YYY; in original form ZZZ}

\pubyear{2021}

\begin{document}
\label{firstpage}
\pagerange{\pageref{firstpage}--\pageref{lastpage}}
\maketitle

\begin{abstract}
Certain types of silicon carbide (SiC) grains, e.g., SiC-X grains, and low density (LD) graphites are C-rich presolar grains that are thought to have condensed in the ejecta of core-collapse supernovae (CCSNe). In this work we compare C, N, Al, Si, and Ti isotopic abundances measured in presolar grains with the predictions of 21 CCSN models. The impact of a range of SN explosion energies is considered, with the high energy models favouring the formation of a C/Si zone enriched in $^{12}$C, $^{28}$Si, and $^{44}$Ti. Eighteen of the 21 models have H ingested into the He-shell and different abundances of H remaining from such H-ingestion. CCSN models with intermediate to low energy (that do not develop a C/Si zone) cannot reproduce the $^{28}$Si and $^{44}$Ti isotopic abundances in grains without assuming mixing with O-rich CCSN ejecta. The most $^{28}$Si-rich grains are reproduced by energetic models when material from the C/Si zone is mixed with surrounding C-rich material, and the observed trends of the $^{44}$Ti/$^{48}$Ti and $^{49}$Ti/$^{48}$Ti ratios are consistent with the C-rich C/Si zone. For the models with H-ingestion, high and intermediate explosion energies allow the production of enough $^{26}$Al to reproduce the  $^{26}$Al/$^{27}$Al measurements of most SiC-X and LD graphites. In both cases, the highest $^{26}$Al/$^{27}$Al ratio is obtained with H still present at $X_H \approx 0.0024$ in He-shell material when the SN shock is passing. The existence of H in the former convective He-shell points to late H-ingestion events in the last days before massive stars explode as a supernova.
\end{abstract}


%%%%%%%%%%%%%%%%%%%%%%%%%%%%%%%%%%%%%%%%%%%%%%%%%%

%%%%%%%%%%%%%%%%% BODY OF PAPER %%%%%%%%%%%%%%%%%%

\section{Introduction} %%%%%%%%%%%%%%%%%%%%%%%%%%%%%%%%%%%%%%%%%%%%%%%%%
\label{sec:intro}

Presolar grains were produced by different types of stars nearby the Sun before its formation about 5 Gyr ago. These grains were important ingredients for the first aggregates of solid materials from which the parent bodies of primitive meteorites formed inside the protoplanetary disk around the young Sun. These meteorites have been recovered on Earth and can therefore be analysed in laboratories. The C-rich grains can be isolated from solar material by dissolving the parent meteorite in acids to leave the resilient grains behind \citep[e.g.][]{Amari1994}. The isotopic ratios and abundances of presolar grains can be measured in laboratories providing a powerful diagnostic for theoretical stellar simulations. 

Few isotopic ratios are observable in stellar spectra, and they suffer from much larger uncertainties compared to data obtained from presolar grains. We refer to e.g.,  \cite{carretta:00,spite:06} for C isotopic ratios, to \cite{hedrosa:13} for N, to \cite{yong:03,carlos:18} for Mg and to \cite{chavez:09} for Ti. %There are numerous types of presolar grains, each originating from different stellar sources and identified by the isotopic abundance measurements \citep[and references therein]{Zinner2014}. 

\begin{figure} %% C vs N (grains only) %%%%%%%%%
    \centering
    \includegraphics[width=0.47\textwidth]{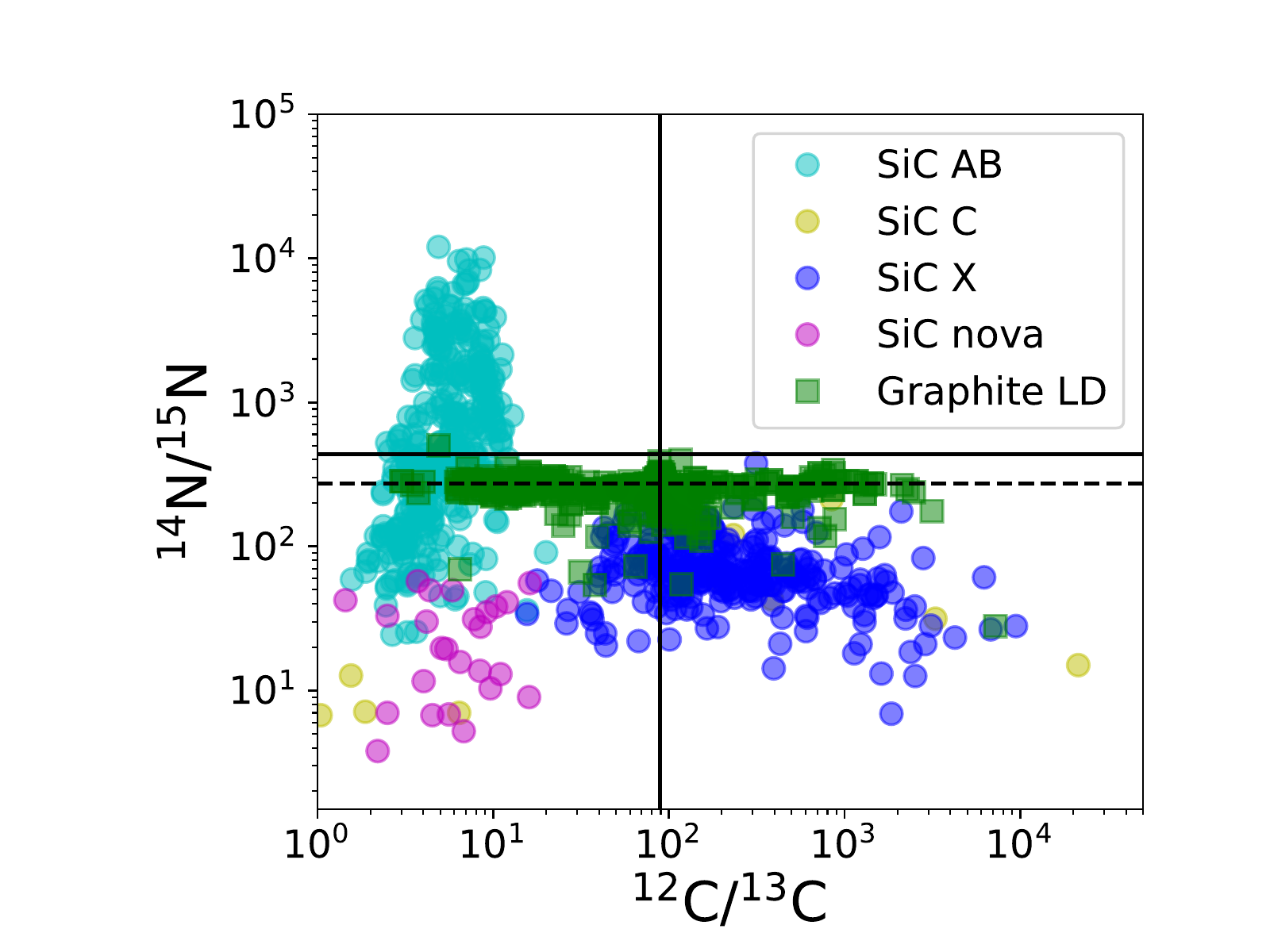}
    \caption{$^{12}$C/$^{13}$C and $^{14}$N/$^{15}$N isotopic ratios of individual presolar grains. SiC sub-types of possible SN origin (AB, C, X, and nova) are shown as well as LD graphites. The solar carbon and nitrogen ratio lines are plotted as solid lines as well as the terrestrial nitrogen ratio as a dashed line ($^{14}$N/$^{15}$N=272). The solar $^{14}$N/$^{15}$N ratio was taken from \citet{Marty2011}. The SiC and LD graphite grain data were taken from the Washington University Presolar Grain Database \citep{Stephan2020,Hynes2009}.}
    \label{fig:SiCgrains}
\end{figure}

Certain types of C-rich grains, e.g., SiC-X grains and low density (LD) graphites, formed in the ejecta of core collapse supernovae \citep[CCSNe; the collapse of a massive star with initial mass $\geq$ 9 M$_\odot$, e.g.][]{Woosley2002,Nomoto2013,Sukhbold2016,Limongi2018,curtis:19}, identified because their isotopic abundances match the signatures of these events. Such signatures of CCSNe include $^{15}$N-excesses relative to solar\footnote{Unless otherwise stated to be absolute, all excesses and deficiencies of isotope abundances are discussed relative to solar ratios.}, strong Si isotope anomalies, and evidence for the presence of the radioactive isotope $^{44}$Ti at the time of grain formation. 
Discovered by \citet{Bernatowicz1987}, SiC grains are a C-rich type of presolar grain that are typically observed at concentrations of $<50$ppm (but up to $150$ppm) and diameters of $<0.5 \umu$m and up to $30 \umu$m \citep[and references therein]{davis:11,Zinner2014,nittler:16}. SiC and graphite grains are believed to only form in C-rich environments, i.e. where C/O $> 1$ (by number fraction). The strength of the CO molecular bond is such that no other C-rich molecules could efficiently form \citep{Ebel2001}.
There are several sub-types of presolar SiC grains (Mainstream, AB, C, X, Y, Z, and putative nova; which are shown in Figure \ref{fig:SiCgrains} with the exception of mainstream, Y, and Z grains) which are characterised by specific isotopic ratios. Among other ratios, the most used to distinguish between different presolar grain types are $^{12}$C/$^{13}$C, $^{14}$N/$^{15}$N, the \footnote{$\delta$(ratio) = (grain ratio/solar ratio - 1) $\times 1000$}delta ratio $\delta$ $^{29,30}$Si and $^{26}$Al/$^{27}$Al %, among others 
\citep{Zinner2014}. 

SiC-X grains \citep[about $1\%$ of all SiC grains, e.g.][]{Zinner2014} have a relative excess of $^{28}$Si to $^{29,30}$Si compared to solar ratios. Most have a $^{12}$C, $^{15}$N excess, while some have a $^{13}$C, $^{15}$N excess compared to solar. The $^{26}$Mg and $^{44}$Ca excesses are due to the condensation of unstable $^{26}$Al and $^{44}$Ti respectively in the grains \citep[e.g.][]{Besmehn2003}. The majority of SiC-X grains also show an excess of $^{49,50}$Ti relative to $^{48}$Ti and 10-20$\%$ show a large excess of $^{44}$Ca \citep{Nittler1996,Besmehn2003,Lin2010}. SiC-X are known to have originated from CCSNe, primarily due to their excesses of $^{28}$Si and extinct $^{44}$Ti \citep[e.g.][]{Nittler1996}.
Evidence for initial $^{44}$Ti within at least one of the few SiC grains classified as nova grains led to the conclusion that they had formed by CCSNe as $^{44}$Ti production is a notable signature of %SNe and SN grains 
CCSN nucleosynthesis
\citep{Nittler2005}. SiC type-AB grains \citep[$\sim 4-5\%$ of all SiC grains;][]{Amari2001} are believed to have multiple sources including CCSNe \citep{Liu2017,Hoppe2019}. Putative nova grains, and $^{15}$N-rich and possibly also $^{15}$N-depleted AB grains are consistent with the H-ingestion models discussed in this work, in \citet{Pignatari2015} and in \citet{Liu2016}.

Low density graphites \citep{Amari1995} are another type of C-rich grain, showing a large $^{12}$C/$^{13}$C ratio range. Excesses and deficits of $^{28}$Si relative to $^{29,30}$Si are both observed, as are excesses in $^{26}$Mg and $^{44}$Ca. With few exceptions, the $^{14}$N/$^{15}$N ratio of LD graphites is mostly consistent at about 300, suggesting heavy contamination from terrestrial nitrogen \citep[e.g.][]{Amari1993,Hoppe1995}. Most LD graphites are believed to originate in CCSNe due to their large range of C isotopic ratios and excesses in signature isotopes such as $^{15}$N, $^{18}$O, $^{26}$Mg, $^{28}$Si, $^{44}$Ca, and $^{49,50}$Ti \citep{Travaglio1999}.

Isotopic signatures such as those found in presolar grains are formed in very specific regions of CCSNe. Classically, the main C-rich region is the He/C zone (the He-shell of the ejecta) above the O-rich O/C zone \citep[][]{Meyer1995}. However, \citet{Pignatari2013b} introduced a new set of models showing a new C-rich zone which was able to explain most observed abundances of C-rich grains. As the shockwave of the CCSN reaches the bottom of the He/C zone, its density and temperature increase dramatically. These conditions efficiently activate explosive He-burning. However, at T > $3.5 \times 10^8$ K the $^{12}$C($\alpha$,$\gamma$)$^{16}$O reaction is up to orders of magnitudes weaker relative to the subsequent reactions in the $\alpha$-capture chain, starting from the reaction $^{16}$O($\alpha$,$\gamma$)$^{20}$Ne.
This leads to a $^{12}$C- and $^{28}$Si-rich, $^{16}$O-deficient region in the deepest He/C shell material, defined by \citet{Pignatari2013b} as the C/Si zone. Additionally, non-homogeneous H-ingestion into the He/C zone prior to the explosions results in the variable production of $^{13}$C, $^{15}$N, and $^{26}$Al by explosive H-burning deep in the C-rich He-shell, nearby the C/Si zone. H-ingestion combined with the extreme conditions of the SN shockwave triggers the production of $^{15}$N via the hot CNO cycle. This allows the isotopic ratios of CCSN grains such as low $^{12}$C/$^{13}$C and $^{15}$N-excesses to be reproduced by local mixing of material \citep[][]{Pignatari2015,Liu2016,Liu2018,Hoppe2018}. 
%It was confirmed by \citet{Liu2016,Liu2018} that H needs to be ingested into the He/C zone during the pre-SN phase, leading to a significant reduction of the $^{12}$C/$^{13}$C ratio in the He/C zone due to $^{13}$C production during the SN explosions.

Classically, the explosive $\alpha$-capture at the bottom of the former convective He-shell activates a neutron burst peaking up to $10^{18-20}$ neutrons cm$^{-3}$, produced by the $^{22}$Ne$(\alpha,n)^{25}$Mg reaction. This nucleosynthesis process was called the n-process \citep{Blake1976,Thielemann1979}. Much later, its signature was identified in heavy elemental abundances measured in presolar SiC-X grains such as large excesses of $^{58}$Fe, $^{88}$Sr, $^{96}$Zr, $^{95,97}$Mo, and $^{138}$Ba \citep[e.g.][]{Meyer2000,Pellin2006,Pignatari2018} and in the production of the radioactive isotope $^{32}$Si \citep{Pignatari2015}. However, this scenario has been discounted as a way to explain the r-process because it could not reproduce the r-process abundance pattern in the Solar System \citep{Blake1981}. The presence of H in the He-shell during the SN explosion may lead to a suppression of the n process, where it becomes more probable to destroy $^{22}$Ne by proton capture than by the $^{22}$Ne$(\alpha,n)^{25}$Mg reaction \citep{Pignatari2015}. The production of $^{44}$Ti and $^{40}$Ca in the C/Si zone along with the possible strong depletion of $^{40}$Ca by neutron capture predicts abundances consistent with the observation of large $^{44}$Ca/$^{40}$Ca in the C-rich grains discussed \citep{Pignatari2013b}.
It is still matter of debate if in presolar grains the abundance signatures of H-ingestion coexist with the neutron-capture signature of the n-process. Indeed, the n-process activation would be compatible with the $^{30}$Si- and $^{32}$S-enhancements measured in in some putative nova grains \citep[][]{Liu2016,Hoppe2018}, and with the $^{32}$Si- and $^{50}$Ti-enhancements found in a sample of SiC-AB grains \citep[][]{Liu2017}. On the other hand, measurements of the Sr, Mo and Ba isotopic abundances in SiC-AB grains seem to be compatible with the pre-explosive s-process abundances in the He shell, and not with the n-process \citep[][]{Liu2018}. In the first case, major asymmetries triggered by the H ingestion in the pre-explosive He shell structure would be required in order to explain for the same single grain the signatures of both the n-process and of a late H-ingestion event leaving H behind in the former He shell material. While this may be a natural expectation from multi-dimensional simulations, it cannot be captured by one-dimensional stellar simulations.
Additionally, those abundance signatures of high neutron-density exposures could also be the effect of a local activation of the intermediate neutron-capture process \citep[i-process][]{cowan:77}, following the H ingestion event. The i-process has been proposed as a nucleosynthesis source active in massive stars at low metallicities \citep[e.g.,][]{roederer:16, clarkson:18, banerjee:18}, but it also has been proposed as a source of anomalous signatures in presolar grains for Ti isotopes \citep[HD graphites, ][]{jadhav:13}, Ba isotopes \citep[in some mainstream SiC grains,][]{liu:14} and $^{32}$Si-enrichments \citep[in SiC AB grains,][]{fujiya:13}. In order to verify all of these different scenarios, new abundance observations are required for more presolar grains of different types. 

\begin{figure} % Ejecta layers %%%%%%%%%%%%55
    \centering
    \includegraphics[width=0.47\textwidth]{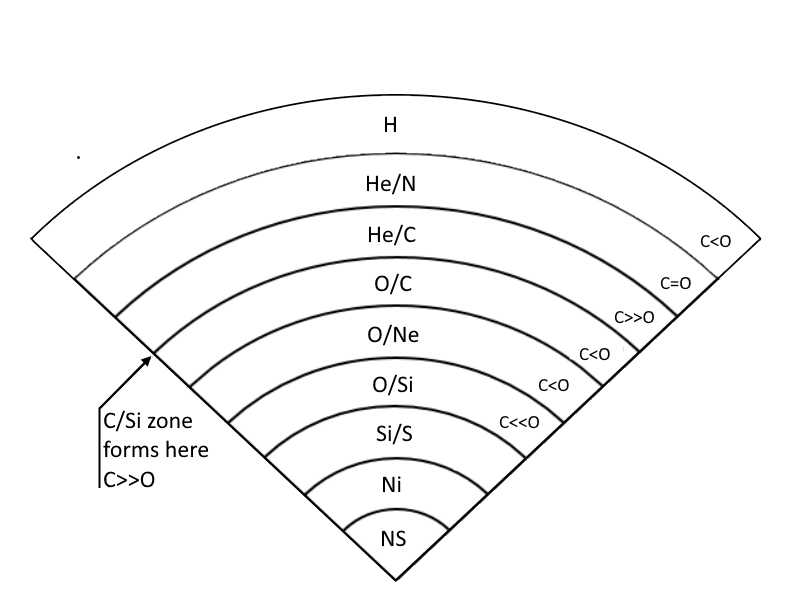}
    \caption{A schematic of the layered 'onion shell' structure of a CCSN ejecta from a massive star progenitor as described by \citet{Meyer1995} with each layer is labelled with its most abundant elements excepting for the central zone. In this case, NS represents the central neutron star formed after the CCSN explosion. The C/Si zone is also indicated at the bottom of the He/C zone. Relative concentrations of C and O are indicated for each zone, with the exception of the deepest Si/S and Ni zones, where there are essentially no C and O left.}
    \label{fig:onion}
\end{figure}

An alternative scenario for the formation of SiC-X grains is that $^{28}$Si and $^{44}$Ti originate from the Si/S zone, which is the region of the ejecta dominated by explosive O-burning \citep{Meyer1995}. 
Figure \ref{fig:onion} shows a schematic of the layered `shell` structure of a  CCSN ejecta as described by \citet{Meyer1995} with the stellar zones. The Si/S zone is mostly made of $^{28}$Si and  $^{32}$S by O fusion, and $^{44}$Ti is efficiently produced by $\alpha$-capture \citep{Woosley1995}. The high $^{12}$C/$^{13}$C ratios are signatures of He-burning and high $^{26}$Al/$^{27}$Al ratios can be produced in the He/N zone by H-burning. However, these isotopic ratios are produced in different, non-neighbouring layers undergoing different phases of burning prior to the SN explosion \citep[e.g.][]{Woosley1995,Rauscher2002}. This requires deep, inhomogeneous mixing of these layers in the CCSN ejecta while maintaining C>O. For instance, Si and Ti signatures may need contributions from the Ni, O/Si, and Si/S zones where complete or partial Si-burning and O-burning occur \citep[e.g.,][]{Travaglio1999,xu:15}. The He/N and He/C zones where H- and incomplete He-burning occurs are also required. This is where the C>O condition is met for SiC condensation \citep{Lodders1997}. This C>O condition necessitates extremely limited addition of material from the intermediate O-rich layers \citep{Travaglio1999}. Three-dimensional hydrodynamic models predict mixing in the ejecta carried out by Rayleigh-Taylor instabilities \citep{Hammer2010}. It remains to be seen if these instabilities allow the inner regions to penetrate (while minimising the addition of material from) the O-rich regions and mix with those rich in C. Astronomical observations of supernovae remnants have shown extensive mixing within the ejected material \citep{Hughes2000,Kifonidis2003}.  

CCSNe also show a large range of explosion energies, from very faint CCSNe with few $10^{50}$erg to hypernovae, up to $5 \times 10^{51}$erg \citep[e.g.,][]{Nomoto2013}.
\citet{Travaglio1999} demonstrated that it was possible to reproduce most isotopic abundances with these models %while assuming C>O 
but also highlighted their shortcomings such as the lack of $^{15}$N and the need for the C-rich condition. 
To overcome this need for mixing between non-neighbouring layers, \citet{Clayton1999} and following works suggested the possibility of condensing C-rich grains in O-rich material. However, this remains controversial as \citet{Ebel2001} concluded that only nanometre-sized grains such as nano-diamonds may form in this turbulent O>C environment and is therefore unrealistic for SiC formation. It was also shown by \citet{Lin2010} that the isotopic ratios of SiC grains of SN origin are inconsistent with the predictions for O-rich material. On the other hand, the advantage of the formation of the C/Si zone is to have isotopes such as $^{28}$Si and $^{44}$Ti made in a C-rich environment, in the vicinity of C-rich He shell layers.

In this work, we compare the abundance predictions of 21 CCSN models with the observed isotopic ratios of C, N, Al, Si and Ti of different types of C-rich presolar grains with established or possible CCSN origin. The structure of the paper is as follows. Section 2 describes the stellar models. In Section 3, the theoretic predictions are compared with observed grain abundances. Section 4 summarises the results.

\section{Theoretical Stellar Models} %%%%%%%%%%%%%%%%%%%%%%%%%%%%%%%%%%%%%%%%%%
\label{sec:models}

In this study, 4 sets of CCSN models are used: two progenitor masses, of 15 and 25 M$_\odot$, both with and without the ingestion of H into the He/C zone in the pre-SN phase. All of the models have a metallicity scaled to Z = 0.02 from the solar abundances by \cite{grevesse:93}. A summary of their main features and specifications can be found in Tables \ref{tab:models1} and \ref{tab:models2}. Detailed information for the models are published in \cite{Pignatari2015} and \cite{Pignatari2016}.

The stellar progenitors were all calculated using the code GENEC \citep[][]{eggenberger:08}. Rotation and magnetic fields were not included. Mass-loss rates during the Main Sequence phase (log\,T$_{\rm eff}$ $>$ 3.9, where T$_{\rm eff}$ is the effective temperature) were taken from \cite{vink:01}. For 3.7 $<$ log\,T$_{\rm eff}$ $<$ 3.9, \cite{dejager:88} prescriptions were used, and for more advanced stages and log\,T$_{\rm eff}$ $<$ 3.7, a scaling law with the stellar luminosity was adopted \citep[][]{Pignatari2016}. Consistently with typical GENEC stellar simulations, the nuclear network to follow energy generation is made of isotopes included explicitly ($^1$H, $^3$He, $^4$He, $^{12}$C, $^{13}$C, $^{14}$N, $^{15}$N, $^{16}$O, $^{17}$O, $^{18}$O, $^{20}$Ne, $^{22}$Ne, $^{24}$Mg, $^{25}$Mg, $^{26}$Mg, $^{28}$Si, $^{32}$S, $^{36}$Ar, $^{40}$Ca, $^{44}$Ti, $^{48}$Cr, $^{52}$Fe, $^{56}$Ni) and implicitly in chain reactions to correctly follow the energy generation up to Si burning \citep[e.g.,][]{hirschi:07}. 

For the CCSN explosion, the mass cut above which the material is ejected (i.e., the mass of the final compact remnant) is given by the analytical formula of \cite{Fryer2012}, depending on the initial mass and metallicity of the progenitor. The amount of matter that falls back after the launch of the explosion is also taken into account. For the stellar layers ejected, the main information for the CCSN nucleosynthesis are given by a semi-analytic description for the shock heating and following temperature and density evolution in the different layers. The shock velocity through the stellar structure is defined by the Sedov blastwave solution \citep[][]{sedov:46}. The peak post-shock temperature and density calculated for each zone of the CCSN ejecta are followed by an adiabatic exponential decay for few seconds, when only $\beta$-decay nuclear reactions remain active from radioactive unstable isotopes \citep[][]{Pignatari2016}.    

The thermodynamic and structural information from the stellar models and from the CCSN explosions are used as input for the post-processing nucleosynthesis calculations which generates complete stellar yields. For these models, we used the parallel multi-zone code mppnp from the NuGrid post-processing simulation tool framework. The nuclear network is dynamic, increasing up to more than 5200 isotopes and 74000 reactions depending on the needs from the nucleosynthesis conditions found. The detailed description of the post-processing code, the simulation setup and the nuclear reaction network is given in \cite{Pignatari2016}. 

Explosive He burning conditions are particularly relevant for our study. The triple-$\alpha$ and $^{12}$C($\alpha$,$\gamma$)$^{16}$O reaction rates are from \cite{fynbo:05} and \cite{kunz:02}, respectively. The $^{22}$Ne($\alpha$,n)$^{25}$Mg reaction rate is from \cite{jaeger:01}. More recent rates are becoming available for this reaction \citep[e.g.,][]{talwar:16,ota:21,adsley:21}, which is the dominant source of neutrons in explosive He burning \citep[][]{Thielemann1979, Meyer2000, Pignatari2018}. However, nuclear uncertainties should have a limited impact at the relevant temperatures for this study, of the order of $10^9$ Kelvin or higher. Neutron capture rates are from the KADoNIS 0.3 compilation where available \citep[][]{dillmann:06}, otherwise for unstable isotopes the theoretical rates adopted were from the %Basel REACLIB database, revision 20090121 \citep[][]{rauscher:00}. 
JINA REACLIB version 1.1 \citep[e.g.,][]{cyburt:10}. Available stellar $\beta$-decay rates are from \cite{fuller:85} and \cite{oda:94} for light isotopes, and for the iron group isotopes and beyond we used \cite{langanke:00} and \cite{goriely:99}.

For this work, we use three 15M$_\odot$ models by \cite{Pignatari2013b} and \cite{Pignatari2016}.
The explosion simulations include the fallback prescription by \citet{Fryer2012} and use the recommended initial shock velocity for one case, while two further cases with the shock velocity divided by two and four have also been calculated (models 15d, 15r2 and 15r4, respectively). 
For model 15d, the shock velocity of material that does not fall back %the shock velocity beyond fallback 
is $2 \times 10^9$cm s$^{-1}$, corresponding to an explosion energy of 3-$5 \times 10^{51}$erg. Models 15r2 and 15r4 correspond to SN explosion energies of 1-$2 \times 10^{51}$erg and less than $10^{51}$erg, respectively \citep[][]{Pignatari2013b,Pignatari2016}. Note that with the  semi-analytic approach used for our models, the SN shock is driven off the proto-neutron star based on the mass cut derived from \cite{Fryer2012}. With this approach, the initial shock velocity becomes a crucial parameter, directly related to the temperature and density peaks in different stellar layers during the CCSN explosion \citep[e.g.,][]{Pignatari2016,ritter:18}. However, shock velocities cannot be mapped directly to CCSN explosion energies. Therefore, for each model we provide an indicative range of CCSN energies corresponding to the given initial shock velocities, instead of a single explosion energy.

Our approach for the CCSN models is different from other studies in the literature, where several methods were adopted to produce large scale one-dimensional nucleosynthesis and abundance yield calculations. Different engines have been used to artificially drive the explosion, among others a piston \citep[][]{Woosley1995, Rauscher2002, sieverding:18}, a thermal bomb \citep[e.g., ][]{thielemann:96, Nomoto2013}, the injection of thermal energy \citep[e.g.,][]{limongi:03}, or more recently effective neutrino-driven models \citep[][]{perego:15, Sukhbold2016, curtis:19}. While the greatest impact of the methods used to simulate the CCSN explosion are expected to be seen in the most internal layers of the CCSN ejecta where the iron group elements are made, the more external regions of the CCSN ejecta relevant for this work (the former C-rich He shell) are not much affected by the adopted approach to simulate the explosion. A much greater impact is instead due to the evolution of the stellar progenitor and to the uncertainties in the progenitor structure \citep[e.g.,][]{Pignatari2015}.

In the present simulation framework, for the most energetic CCSN models described above (15d and 15r2), the temperature rises from $3.1 \times 10^8$K in the pre-SN stage to a peak of up to about $2.0 \times 10^9$K %at the bottom of 
in the forming C/Si zone \citep{Pignatari2013b}. Note that for these models in the former O/C zone and at the bottom of the C/Si zone there is still a small region where O is destroyed but there is not enough $^{4}$He to make $^{28}$Si. Here the temperature rises up to about $2.2 \times 10^9$K (see Table \ref{tab:models1}). In the region of the former He shell relevant for explosive nucleosynthesis, the SN peak temperature in model 15r4 is about 30-40\% lower than in 15d, while 15r2 and 15d are the same.

The 25d model is calculated for a 25M$_\odot$ progenitor star. Since no substantial $^{56}$Ni is ejected by the explosion, we may classify 25d as a faint SN model \citep[][]{Pignatari2016}. 
After core C-burning ends in the progenitor star, the convective He-shell becomes unstable and ingests H into it from above the He-rich region. The beginning of core O-burning deactivates the convective zone completely until the SN explosion, leaving the He-rich shell with a H abundance of about $1.2\%$. 
With 25d, \citet{Pignatari2015} introduced 25d-H5, 25d-H10, 25d-H20, 25d-H50, and 25d-H500 (25d model set). These models are the same as 25d in all respects except the H abundance in the He-shell is reduced by a factor of 5, 10, 20, 50, and 500 respectively.

\begin{figure} %% 25T-H and H20 abundances %%%%%%%
    \centering
    \includegraphics[width=0.48\textwidth]{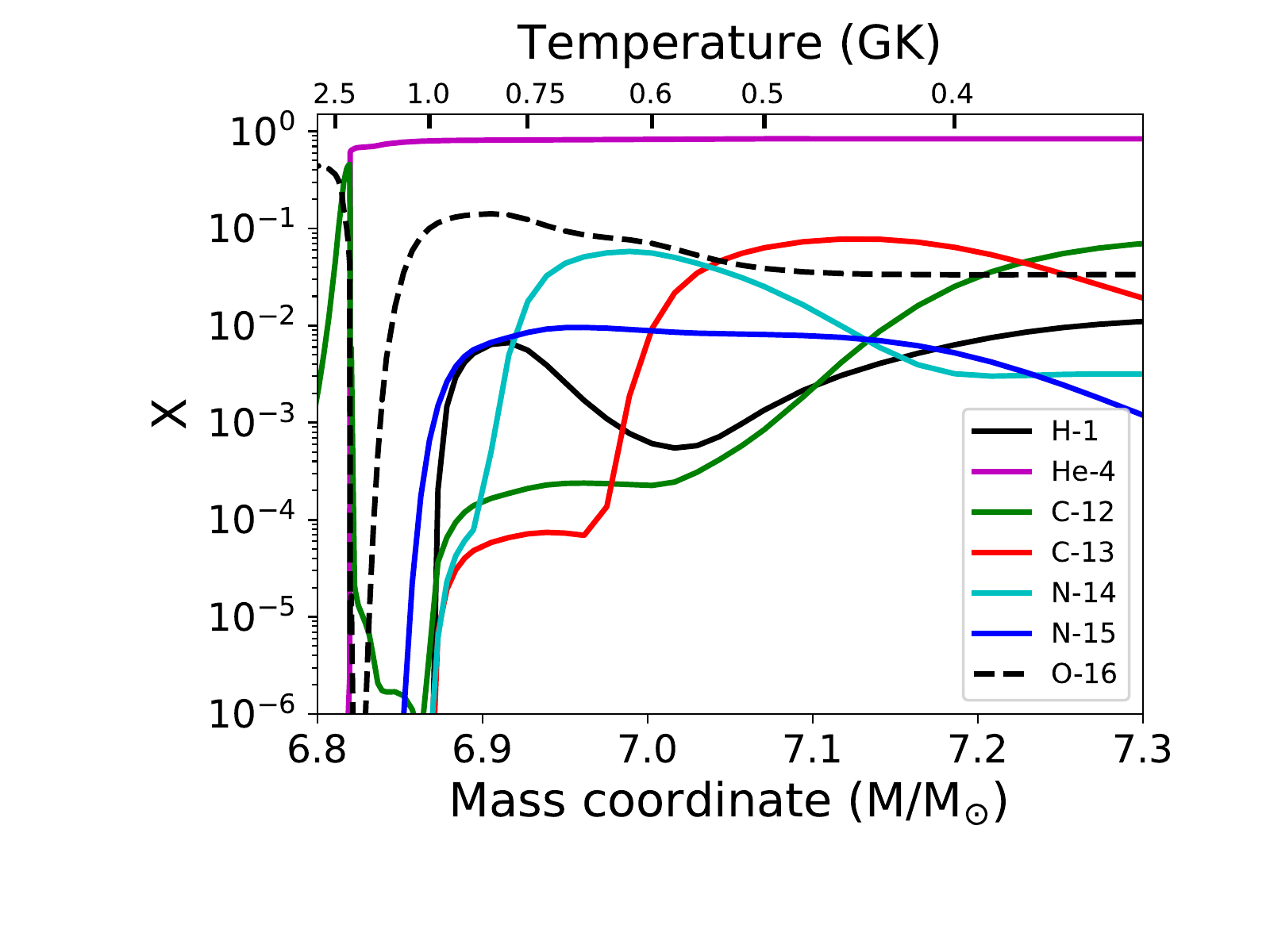}
    \includegraphics[width=0.48\textwidth]{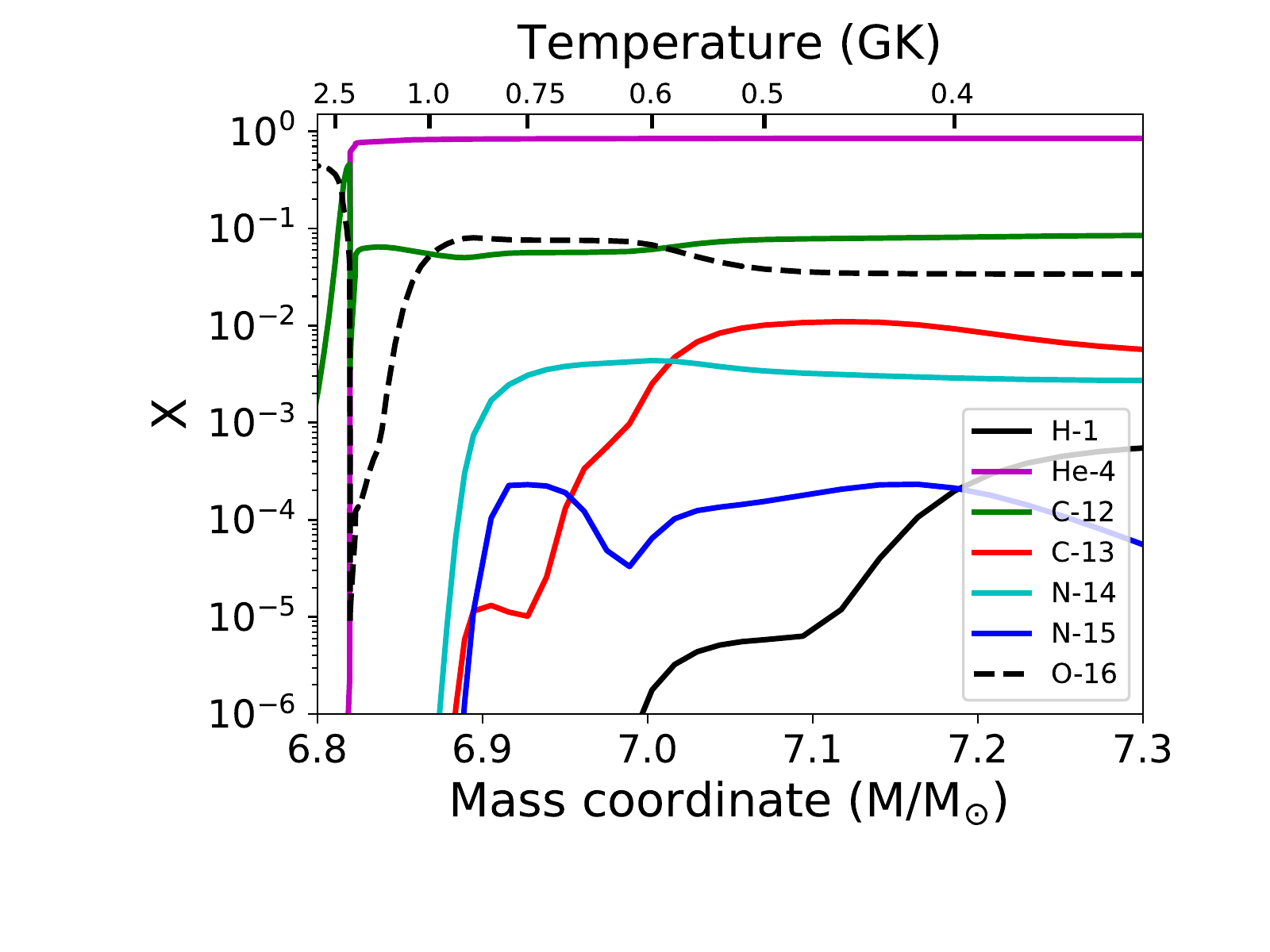}
    \caption{Isotopic abundances (mass fractions) in the lower He-shell ejecta of the 25T-H and 25T-H20  models (upper and lower panels respectively) immediately after the CCSN explosion. The abundance profiles for H, $^{4}$He, $^{12,13}$C, $^{14,15}$N and $^{16}$O are shown, taking into account their radiogenic contributions. The secondary x-axis on the top is showing the scale of the CCSN shock peak temperatures in GK (10$^9$ K units) at different mass coordinates. }
    \label{fig:25T_abund}
\end{figure}

\citet{Pignatari2015} also introduced the 25T model set where the temperature and density of 25d are artificially increased to reproduce the explosive conditions of a 15M$_\odot$ star in the He shell layers (15d model, included in this study). The bottom of the C/Si zone in the 15d model reaches a peak temperature 3.3 times higher than in the 25d model ($2.3 \times 10^9$K rather than $0.7 \times 10^9$K), with the 25T model reaching $2.2 \times 10^9$K. The peak density is also 100 times higher. The same range of pre-explosion He-shell concentrations of H are examined here; 25T-H, 25T-H5, 25T-H10, 25T-H20, 25T-H50, and 25T-H500. As an example, the abundance profiles in the He-shell ejecta of 25T-H and 25T-H20 are shown in Figure \ref{fig:25T_abund}.  
Relevant differences are the stronger $^{12}$C depletion in 25T-H compared to 25T-H20, partially compensated by the production of $^{13}$C. The pre-SN $^{14}$N and $^{15}$N abundances are both increased by the SN explosion in model 25T-H, by up to an order of magnitude and two orders of magnitudes respectively in some of the He shell zones shown in the figure. On the other hand, in the 25T-H20 model the N ejecta will be more representative of the pre-SN abundances, defined by the specific H ingestion event \citep[][]{Pignatari2015}.
In Figure \ref{fig:25T_abund}, for both the two models we can see the formation of the so-called O/Nova zone in the ejected CCSN abundances just above the bottom of the He shell \citep[][]{Pignatari2015}, where oxygen is more abundant than carbon. Indeed, the combined $^{12}$C depletion and $^{16}$O production allows an O-rich region to form from the C-rich pre-explosive He shell. 
In these conditions, the explosive ``Nova-like'' H-burning nucleosynthesis on the previous H-ingestion products provides a nucleosynthesis characterized by high enrichments of e.g., $^{15}$N, $^{25}$Mg and $^{26}$Al. Such anomalous signatures have been studied by several presolar grains studies recently, supporting the existence of the conditions discussed here based on isotopic measurements of both presolar SiC grains \citep[][]{Liu2017, liu:18, Hoppe2019} and presolar silicates \citep[][]{hoppe:21}. For each model, the mass coordinates of the O/Nova zones formed during the CCSN explosions are given in Table ~\ref{tab:models2}.   

In this work we also present a new set of six CCSN models, aiming at refining the study of the impact of different explosion energies in models with H-ingestion. The 25av model set was designed to be an intermediate to the 25d and 25T sets by having its shock peak temperature and peak density at the midpoint of the two. Again, the 25av set has a range of H concentrations as described above. These are labelled as: 25av-H, 25av-H5, 25av-H10, 25av-H20, 25av-H50, and 25av-H500. Note that the temperature at the bottom of the C-rich region of models 25av is similar to the temperatures found in model 15r4 (see Table \ref{tab:models2}).

%A summary of all the models in this work and their specifications can be found in Table \ref{tab:models} for reference.

\begin{table*} %% Summary Table of Models %%%%%%
    % \centering
    % \small % smaller text 
    \begin{tabular}{|c|c|c|c|c|c|c|}
        \hline \hline
        \multicolumn{7}{|c|}{Models without H-ingestion} \\
        \hline \hline
        Model set 15d & \multicolumn{2}{|c|}{15d} & \multicolumn{2}{|c|}{15r2} & \multicolumn{2}{|c|}{15r4} \\
        Shock Peak Temperature [K] & \multicolumn{2}{|c|}{$2.2 \times 10^9$} & \multicolumn{2}{|c|}{$2.2 \times 10^9$} & \multicolumn{2}{|c|}{$1.6 \times 10^9$} \\
        C/Si zone [M$_\odot$] & \multicolumn{2}{|c|}{$2.945-2.975$} & \multicolumn{2}{|c|}{$2.945-2.975$} & \multicolumn{2}{|c|}{$2.945-2.975$}\\
        He-shell [M$_\odot$] & \multicolumn{2}{|c|}{$2.945-3.3$} & \multicolumn{2}{|c|}{$2.945-3.3$} & \multicolumn{2}{|c|}{$2.945-3.2$} \\
        Explosion Energy [erg] & \multicolumn{2}{|c|}{$3-5 \times 10^{51}$} & \multicolumn{2}{|c|}{$1-2 \times 10^{51}$} & \multicolumn{2}{|c|}{$<10^{51}$} \\
        \hline \hline
     \end{tabular}
    \caption{The three 15M$_\odot$ CCSN models used for the work and introduced by \citet{Pignatari2013b} are summarised. They are without H-ingestion. For these models, the table shows the mass coordinates of the He-shell and C/Si zone, the shock peak temperature at the bottom of the deepest C-rich ejecta (i.e., the C/Si zone in the 15d and 15r2 models) as well as the explosion energy range of each model.} 
    \label{tab:models1}
\end{table*}

\begin{table*} %% Summary Table of Models %%%%%%
    % \centering
    % \small % smaller text 
    \begin{tabular}{|c|c|c|c|c|c|c|}
        \hline \hline
        \multicolumn{7}{|c|}{Models with H-ingestion} \\
        \hline
         X$_H$ in He-shell & $1.2\%$ & $0.24\%$ & $0.12\%$ & $0.06\%$ & $0.024\%$ & $0.0024\%$ \\
        before SN shock \\
        \hline \hline
        Model set 25T & 25T-H & 25T-H5 & 25T-H10 & 25T-H20 & 25T-H50 & 25T-H500 \\
        Shock Peak Temperature [K] & \multicolumn{6}{|c|}{$2.3 \times 10^9$} \\
        C/Si zone [M$_\odot$] & \multicolumn{6}{|c|}{$6.815-6.820$} \\
        O/nova zone [M$_\odot$] & $6.83-7.04$ & $6.84-7.02$ & $6.86-7.00$ & $6.87-6.96$ & ... & ... \\
        He-shell [M$_\odot$] & \multicolumn{6}{|c|}{$6.81-9.23$} \\
        \hline
        Model set 25av & 25av-H & 25av-H5 & 25av-H10 & 25av-H20 & 25av-H50 & 25av-H500 \\
        Shock Peak Temperature [K] & \multicolumn{6}{|c|}{$1.5 \times 10^9$} \\
        C/Si zone [M$_\odot$] & \multicolumn{6}{|c|}{...} \\
        O/nova zone [M$_\odot$] & $6.82-6.87$ & $6.82-6.85$ & $6.82-6.85$ & $6.82-6.84$ & ... & ... \\
        He-shell [M$_\odot$] & \multicolumn{6}{|c|}{$6.82-9.23$} \\
         \hline
        Model set 25d & 25d-H & 25d-H5 & 25d-H10 & 25d-H20 & 25d-H50 & 25d-H500 \\
        Shock Peak Temperature [K] & \multicolumn{6}{|c|}{$0.7 \times 10^9$} \\
        C/Si zone [M$_\odot$] & \multicolumn{6}{|c|}{...} \\
        O/nova zone [M$_\odot$] & \multicolumn{6}{|c|}{...} \\
        He-shell [M$_\odot$] & \multicolumn{6}{|c|}{$6.82-9.23$} \\
       \hline \hline
    \end{tabular}
    \caption{The 18 CCSN models used for the work and with H-ingestion are summarised as follows. 
    The 25M$_\odot$ models have H-ingestion into the He/C in the pre-SN phase. Model 25d was fully published by \citet{Pignatari2016}. The rest of the 25d set and the 25T set were introduced by \citet{Pignatari2015}. The 25av set is introduced in this work. For each H-ingestion model, the table shows the shock peak temperature at the bottom of the deepest C-rich ejecta (i.e., the C/Si zone in the 15d and 25T models, a C-rich region with mild O depletion triggered by explosive He-burning for the 25av models and the He/C zone for the 25d models), mass coordinate of the He-shell and O-rich O/nova zone, and the abundance of H remaining in the He-shell from ingestion before the SN shock.}
    \label{tab:models2}
\end{table*}

\section{Comparison with presolar grains} %%%%%%%%%%%%%%%%%%%%%%%%%%%%%%%%%%%%%%
\label{sec:results}

In this section, we compare the stellar models to the grain data.
We will only consider the abundance profiles in C-rich regions of the SN ejecta for the comparison, i.e., the He/C zone and the C/Si zone (if the SN conditions allow its formation). For the models affected by H ingestion, we will also consider the O/nova zone as discussed by \cite{Pignatari2015}. Indeed, although it is O-rich, its position between the C/Si and the He/C zones makes it reasonable to expect that local mixing of stellar material would still result in C-rich mixtures. 

\begin{figure*} %% C/N plot %%%%%%%%%%%%%
    \centering
    \includegraphics[width=1.0\textwidth]{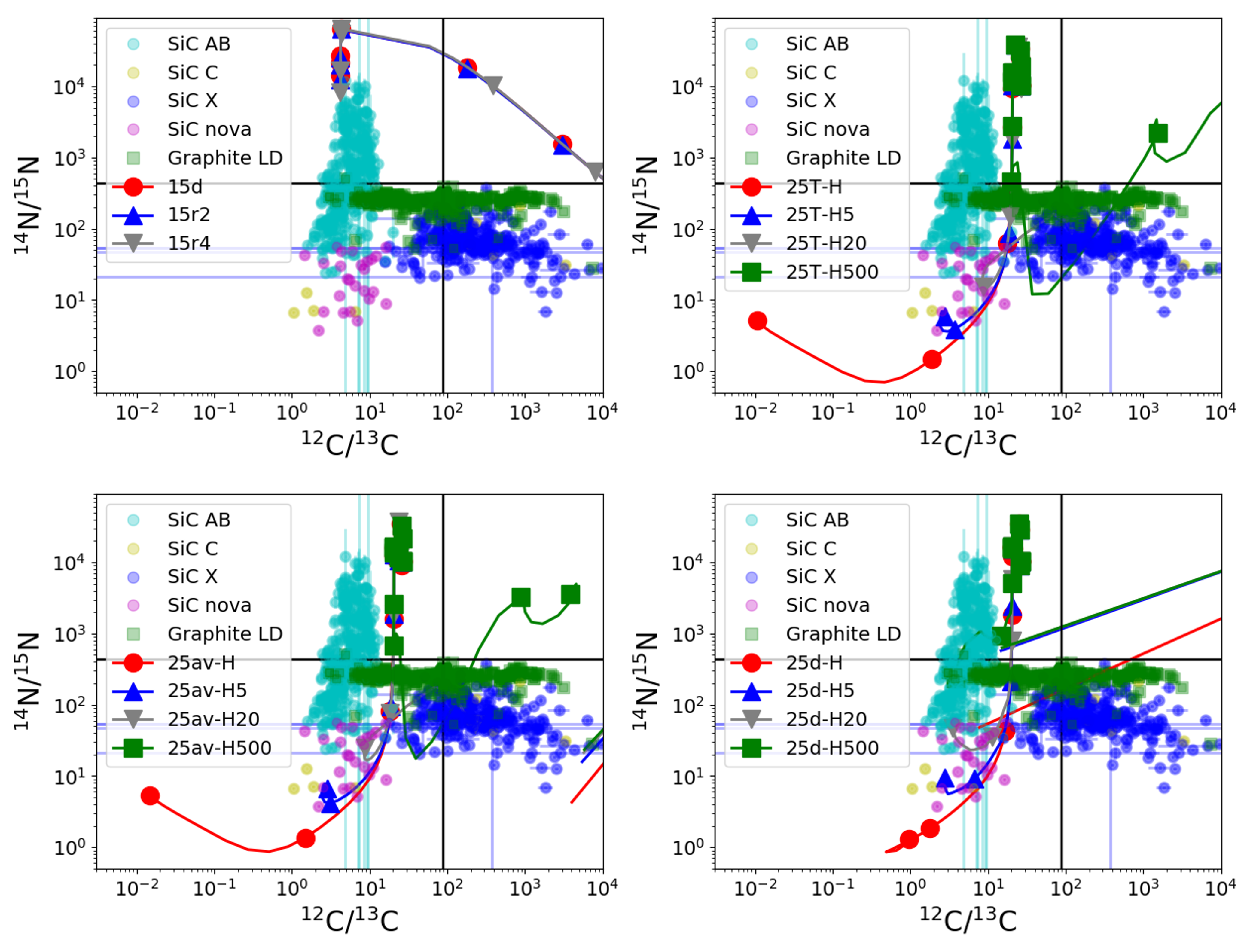}
    \caption{$^{14}$N/$^{15}$N and $^{12}$C/$^{13}$C isotopic ratios as predicted across the He/C zone of four sets of models, compared with the measured data for individual SiC grains (sub-types AB, nova, X, and C) and LD graphites. % Panel 1 (upper left)
    The top left Panel shows the CCSN abundances for the 15d, 15r2, and 15r4 models, exploring different SN energies. %Panel 2 (upper right) 
    The top left Panel shows the 25T set that has a single energy (energetically similar to 15d but has different initial conditions) and explores the H left from ingestion. %Panels 3 and 4 (lower left and right respectively) 
    The bottom left and right Panels show the 25av and 25d sets respectively, which have same parameter space for H but have different energies. The continuous vertical and horizontal black lines indicate the solar ratios. %The numbers in the plots indicate the mass coordinates where the C-rich regions start and end in the ejecta.
    }
    \label{fig:CvsN}
\end{figure*}

In Figure \ref{fig:CvsN}, the C and N isotopic ratios of individual SiC grains and LD graphites are compared to the %predictions 
CCSN abundances predicted from across the He/C zone of the models. The 15d, 15r2, and 15r4 models in %panel 1 (upper left) 
the top left Panel have the same stellar progenitor but have different explosion energies and no H-ingestion. The 25T models in %panel 2 (upper right) 
the top right Panel have the same explosion energy (extremely similar to that of 15d, see \S~\ref{sec:models}) but have different initial conditions. They allow us to explore the impact of different amounts of H remaining from ingestion, ranging from 25T-H (about 1.2\% of H left at the SN shock passage from the previous H-ingestion event) down to 25T-H500 (about 0.0024\%). In the 25T-H500 case, there is no significant explosive H-burning nucleosynthesis taking place, but still has the signatures of H-ingestion such as the production of $^{13}$C and $^{14,15}$N via the CNO-cycle. 

The 25av and 25d model sets are shown in %panels 3 and 4 (lower left and right)
the bottom left and right Panels respectively, covering the same parameter space for H as the 25T set. However, these model sets have different explosion energies (see \S~\ref{sec:models}). % to the 25T set. 
Note that the -H10 and -H50 models are not shown in these figures, since they  appear to follow a linear behaviour with the initial H concentration in between models -H5 and -H500.
As discussed in \citet{Pignatari2013b}, in the 15d model %develops 
a C/Si zone is developed by CCSN nucleosynthesis with large $^{12}$C/$^{13}$C ratios ($10^5$-$10^9$, off the right of the plot).  
%The signatures of these models appear to follow a linear dependence with the initial H concentration, thus the -H10 and -H50 models are not shown in these figures to show the range of predictions while not overcrowding the them.
%, and they show a wide range of $^{14}$N/$^{15}$N ratio of 0.02-500. 
The outer region of the He/C zones and the He/N zone in the 15M$_\odot$ models are affected by H-burning, with a subsolar C ratio of about 4. Assuming different combinations of local mixing between the C/Si zone, the He/C zone and other external CCSN layers, these models would be potentially suited to explain the C and N ratios for the SiC-X grains and LD graphites with C ratios larger than solar \citep{Pignatari2013b}.
For the 25M$_\odot$, H-ingestion models, most of the He/C zone contains CCSN abundances that skirt along the edge of the presolar SiC-AB grains at a $^{12}$C/$^{13}$C ratio of around 20, covering $^{14}$N/$^{15}$N values in the range $10^2$ - $10^4$. In these regions of the ejecta, the $^{12}$C/$^{13}$C ratio is defined by the main properties of the H-ingestion event and it is not affected by the CCSN explosion. Since these models share the same stellar progenitor, they all show the same ratio. On the other hand, we should expect that different H ingestions in different real progenitors could naturally produce a large variation of sub-solar $^{12}$C/$^{13}$C ratios in He-shell layers, possibly even within the same star. Indeed, depending on the timescale and strength of the ingestion the C ratio may change significantly. In order to explore the realistic range of C ratios to be expected from H ingestion in massive stars, multi-dimensional hydrodynamics simulations are needed \citep[e.g.][]{Herwig2014,Woodward2015,clarkson:21}. 
In general, from our models we find that the C-rich He shell in models with H-ingestion carries a much lower C ratio compared to the models without H-ingestion, reaching values much lower than solar. Therefore, these models appeared to be more suited to explain presolar C-rich grains from CCSNe with low C ratios \citep{Pignatari2015}. 

The deepest layers of the He/C zone are strongly affected by explosive H burning, and produce low $^{12}$C/$^{13}$C and $^{14}$N/$^{15}$N ratios that are consistent even with the most extreme abundance signatures measured in putative SiC nova grains \citep{Pignatari2015,Liu2016,Hoppe2018}, and with some degree of local mixing with SiC-AB grains \citep{Liu2017,Hoppe2019}. The differences in the models compared to the upper He/C zone is due to the partial activation of the hot CNO cycle \citep[e.g.,][]{wiescher:10}. Models with more remaining H in the deep He/C zone achieve lower $^{12}$C/$^{13}$C and $^{14}$N/$^{15}$N ratios in the CCSN ejecta due to radiogenic contributions to $^{13}$C and $^{15}$N. %These abundances are consistent with the lowest C and N ratios obtained in SiC-X grains and putative nova grains \citep{Pignatari2015,Liu2016,Hoppe2018}, and with some degree of local mixing with SiC-AB grains \citep{Liu2017,Hoppe2019}. 
For example, 25T-H has the most H remaining from ingestion and develops the lowest $^{12}$C/$^{13}$C and $^{14}$N/$^{15}$N ratios of all the models, while 25T-H500 has no H deep in the He/C zone, so develops high $^{12}$C/$^{13}$C and $^{14}$N/$^{15}$N ratios. The CCSN abundances in the C/Si zone develop %with 
$^{12}$C/$^{13}$C and $^{14}$N/$^{15}$N ratios of about $10^7$ - $10^9$ and 0.01 - $400$ respectively, so with local mixing between the C/Si zone and the deep He/C zone, the 25T-H, 25T-H5, and 25T-H20 models could potentially reproduce all SiC-X grains, including those with $^{12}$C/$^{13}$C ratios higher than solar \citep[e.g.][]{Hoppe2018}.

In %panel 3
the bottom left Panel, despite the much lower SN energy, the 25av models show a similar evolution of the C and N ratios in the CCSN ejecta compared to the 25T set. 
%They differ in that they do not develop a C/Si zone. That region will be more similar to the He/C zone instead, located at the intersection of the O/C zone and the O/Nova zone (for models 25av-H - 25av-H20) or the He/C zone (for models 25av-H50 and 25av-H500, see Table \ref{tab:models}). WE TELL THIS LATER. FROM HERE, IT SEEMS THAT YOU GET A C-RICH REGION STILL JUST BELOW THE O/NOVA ZONE?
The high $^{12}$C/$^{13}$C region to the right of the plot represents the C-rich region at the top of the O/C zone extending from 6.73 to 6.82M$_\odot$. Mixing this region with material in the He/C zone and the intermediate O-rich O/nova zone may explain the large scatter of anomalous ratios measured in presolar grains, still satisfying the constraint of a C-rich mixture. Since the 25av explosions are less energetic than 25T, both less $^{14}$N and $^{15}$N are made at the bottom of the He/C zone, but still showing similar N isotopic ratios overall.
%This explains the formation of similar N isotopic ratios between the two sets of models.

In %panel 4
the bottom right Panel, at the bottom of the He/C zone (6.82M$_\odot$), the models 25d-H5 and 25d-H20 show trends similar to their analogous models at higher energies. However, the lowest C ratio reached by model 25d-H is 0.5, which is not as extreme as the 25T-H and 25av-H models. 
%This further reduction in explosion energy results in 25d-H500 not achieving above-solar $^{12}$C/$^{13}$C and $^{14}$N/$^{15}$N as 25T-H500 and 25av-H500 do, but rather behaves similar to the other 25d model in the deep He/C zone though with higher $^{14}$N/$^{15}$N. 
The SN explosive conditions combined with the lower H concentration allow 25d-H500 and 25d-H20 to provide %This allows 25d-H500 along with 25d-H20 to predict 
$^{12}$C/$^{13}$C and $^{14}$N/$^{15}$N ratios consistent with $^{14}$N-rich SiC-AB grains in the deep He/C zone around 6.82M$_\odot$. Alternatively, \citet{Hoppe2019} suggested adjustments of C and Al isotopic ratios in the 25T models in order to simultaneously match C, N, and Al isotopic ratios of AB grains (both $^{15}$N-rich and $^{15}$N-poor).
%Notice that the 25d, 25av and 25T models all share the same progenitor model. The C isotopic ratios in the outer part of the He/C zone is dominated by the H-ingestion, and it is not affected by the CCSN explosion. Depending on the timescale and strength of the ingestion the C ratio may change significantly. In order to explore the possible range of C ratios, multi-dimensional hydrodynamics simulations are needed \citep[e.g.][]{Herwig2014,Woodward2015}.  Once such a study will be available, it will be possible to see if it is compatible with the C and N isotopic ratios of SiC AB grains.

\begin{figure*} %% dSi plot %%%%%%%%%
    \centering
    \includegraphics[width=1.0\textwidth]{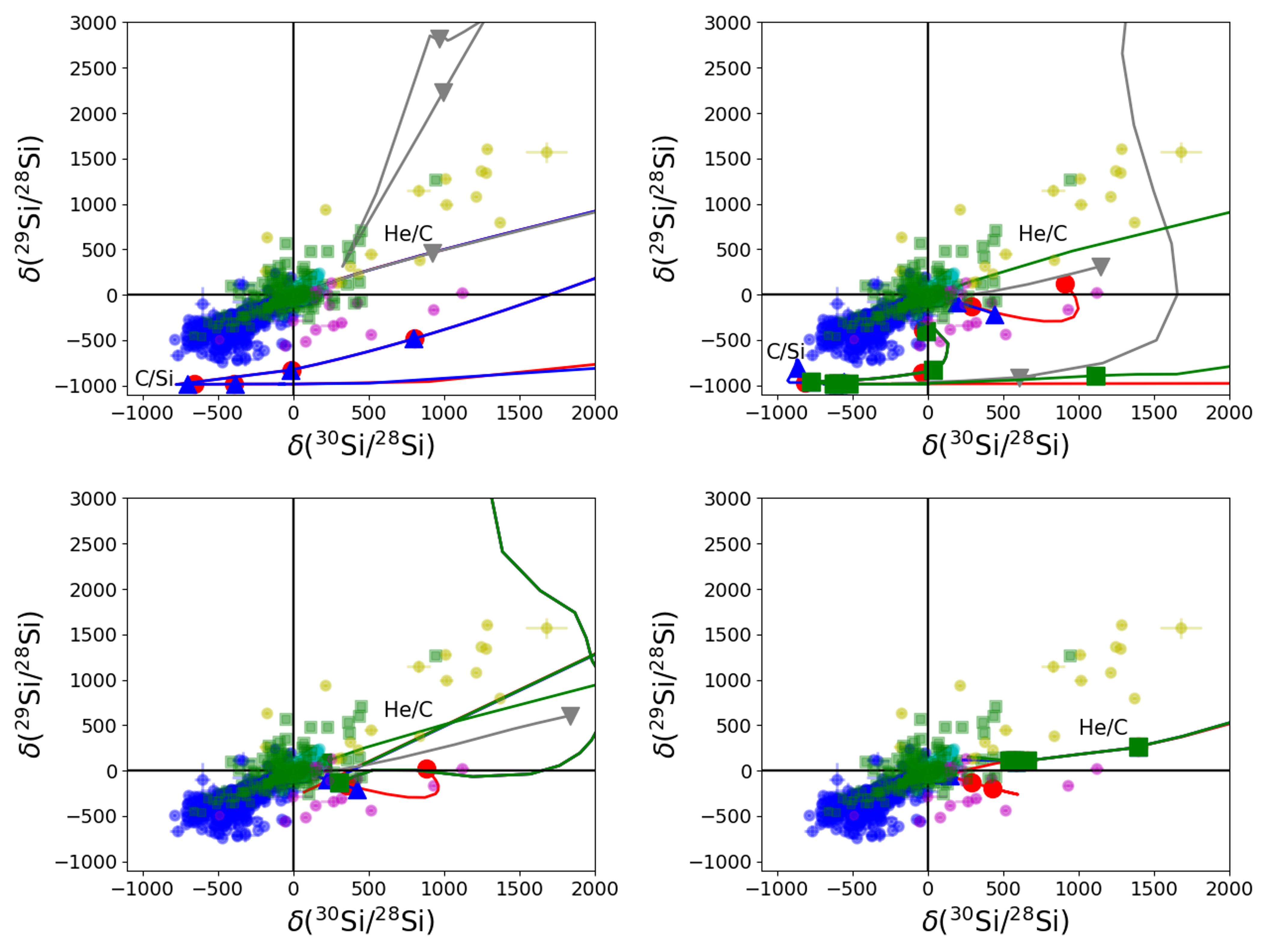}
    \caption{Si isotopic ratios in $\delta$-notation for SiC grains (sub-types AB, nova, X, and C) and LD graphites are compared with the C-rich regions of the models, where $\delta$(ratio) = (grain ratio/solar ratio - 1) $\times 1000$. Abundance ratios measured in grains and CCSN abundance profiles from stellar models are shown in the four panels with the same symbols used in Figure \ref{fig:CvsN}.}
    \label{fig:dSi}
\end{figure*}

In Figure \ref{fig:dSi}, the Si isotopic ratios for SiC grains (sub-types AB, nova, X, and C) and LD graphites are shown with the %predictions 
CCSN abundances predicted from the four model sets. 
In %panel 1
the top left Panel, the 15d and 15r2 models show a large $^{28}$Si-excess in the C/Si zone as shown by low delta values for $^{29}$Si/$^{28}$Si and $^{30}$Si/$^{28}$Si. 
By assuming some mixing with more external stellar material carrying a %$^{29,30}$Si-rich signature 
$^{28}$Si-deficit or a close-to-solar Si isotopic composition, these models can qualitatively reproduce grains which show $^{28}$Si-excess such as SiC-X, a relevant number of LD graphites, and a few nova grains \citep{Pignatari2013b}. On the other hand, in the same region of the ejecta 15r4 does not develop a C/Si zone and the $^{28}$Si-rich signature obtained in 15d and 15r2. %, and observed in these grains \citep{Pignatari2013b}. 
In the intermediate He/C zone of 15d, 15r2, and 15r4, there are strong %$^{29,30}$Si-enrichments  
$^{28}$Si-deficits due to the production of both $^{29}$Si and $^{30}$Si by neutron capture
\citep[this is the neutron-burst signature due to the strong activation of the n-process, see e.g.][]{Meyer2000,pignatari:13c,Pignatari2018}, resulting in higher Si $\delta$-values (up to 18000 and 25000 for $^{30/28}$Si and $^{29/28}$Si respectively) than in the deeper He/C zone. In particular, the more energetic models show a more efficient production of $^{30}$Si compared to $^{29}$Si.
%with the more energetic models showing larger $^{30}$Si-excess than $^{29}$Si. 
This material may be consistent with the %$^{29,30}$Si-excesses 
$^{28}$Si-deficit seen in SiC-C grains that continue off the right of the plotted area, since mixing with material with ratios close-to-solar will result in much less extreme ratios. In the CCSN ejecta, Si isotopic ratios become less extreme in the outer He/C zone, with solar ratios or mild %$^{29,30}$Si excesses
$^{28}$Si-deficit due to the pre-explosive s-process production in the He shell \citep[e.g.,][]{Rauscher2002, Pignatari2016, Sukhbold2016}.

\begin{figure} %% dSi plot zoom
    \centering
    \includegraphics[width=0.45\textwidth]{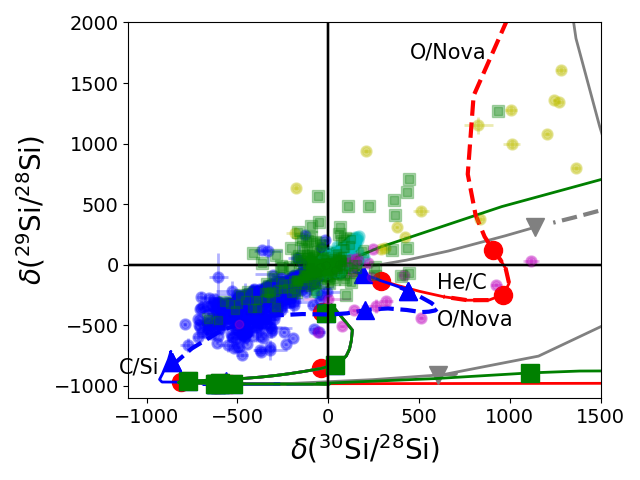}
    \caption{Zoomed version of %panel 2 
    the top right Panel in Figure \ref{fig:dSi}, Only for models 25T-H (red line with circles) and 25T-H5 (blue line with triangles), but showing the isotopic abundances for both C-rich CCSN ejecta (continuous line) and the O-rich O/Nova zone (dashed line). Abundance ratios measured in grains are shown with the same symbols used in Figure \ref{fig:CvsN}.}
    \label{fig:dSi_zoom}
\end{figure}

In %panel 2 
the top right Panel of Figure \ref{fig:dSi}, the 25T models also develop a $^{28}$Si-rich C/Si zone.
%as well as more neutrons made by the alpha-capture on $^{22}$Ne. 
However, with the exception of 25T-H500, the 25T models predict less extreme Si ratios in the He/C zone. Indeed, during the SN explosion proton capture on $^{22}$Ne is in competition with $\alpha$-capture, reducing the activity of the $^{22}$Ne($\alpha$,n)$^{25}$Mg reaction and the strength of the n-process \citep[][]{Pignatari2015,Liu2018}. Therefore, a lower remaining abundance of H from ingestion (e.g. 25T-H500) allows the production of larger %$^{29,30}$Si-excesses. 
$^{28}$Si-deficit. % as there are more neutrons made by the alpha-capture on $^{22}$Ne. 
By comparing with the models in %panel 1
the top left Panel, a novel CCSN nucleosynthesis signature in 25T-H and 25T-H5 is observed at the bottom of the He/C zone. Combined mild %$^{30}$Si-excess and $^{29}$Si-deficiency 
positive $\delta$($^{30}$Si/$^{28}$Si) and negative $\delta$($^{29}$Si/$^{28}$Si) are obtained, to a degree that is directly compatible with measurements in putative nova grains. This is simply due to the explosive H-burning active in that region, favouring the production of $^{30}$Si with respect to $^{29}$Si. The same feature is obtained in simulations for ONe novae \citep[e.g.,][]{amari:01, jose:01, jose:07, denissenkov:14}. In model 25T-H, just at the boundary between the He/C zone and the O/Nova zone, a %$^{29}$Si-excess 
positive $\delta$($^{30}$Si/$^{28}$Si) starts also to be developed. The same signature in nova models can be reached for the most massive ONe novae \citep[see e.g.,][]{amari:01, jose:07a}. However, due to the higher temperatures involved in the O/Nova zone together with the high concentration of H, a much larger %$^{29,30}$Si-excesses 
$^{28}$Si-deficit is achieved. We defer a discussion of the details of O/nova zone nucleosynthesis to a future work. Here we are instead interested in the specific differences obtained in models 25T-H and 25T-H5. These models have the same explosion energy, but different H concentration in the stellar progenitor. In Figure \ref{fig:dSi_zoom}, we show a more zoomed version of the top right Panel %panel 2 
from Figure \ref{fig:dSi}, showing also the evolution of the ejected CCSN abundances in the O/Nova zone. In this region, 25T-H is evolving toward high %$^{29,30}$Si-excesses 
positive $\delta$($^{29}$Si/$^{28}$Si) and $\delta$($^{30}$Si/$^{28}$Si)
up to 35000 and 60000 per mil respectively (outside the plot boundary), while in the C/Si zone the ratios are returning to be consistent with 15d, shown in %Panel 1 
the top left Panel of Figure \ref{fig:dSi}. Model 25T-H5 shows instead a completely different trend. The %$^{29}$Si-deficiency 
negative $\delta$($^{29}$Si/$^{28}$Si) is maintained, while with the higher energies the %low $^{30}$Si/$^{28}$Si 
negative $\delta$($^{30}$Si/$^{28}$Si) is eventually obtained, consistent with the SiC-X observed slope. The SN explosion is clearly not the cause of the difference, since the two models share the same SN prescription. The different trends between 25T-H and 25T-H5 is only due to the different combination of explosive H-burning and explosive He-burning nucleosynthesis in the O/Nova zone, due to the different H concentration. This result is interesting and deserves further study, since it might provide future constraints for stellar models with H ingestion.

In %panel 3 
the bottom left Panel of Figure \ref{fig:dSi}, no C/Si zone is formed by explosive nucleosynthesis due to their lower explosion energy. On the other hand, the nucleosynthesis of Si isotopes in the He/C zone shows several similarities compared to the 25T models with analogous H fuel. Also at these explosion energies the different Si production triggered in 25av-H with respect to 25av-H5 is visible, confirming that the result mostly depend on the H concentration. 
%The 25av models all develop high Si $\delta$-values in the deep He/C zone due to neutron-capture but does not form the $^{30}$Si-excess, $^{29}$Si-deficiency seen in the 15d, 15r2, and 25T models due to the lower energy of 25av.

In %panel 4
the bottom right Panel, the 25d models predict similar high Si $\delta$-values in the He/C zone, with close to solar ratios in the intermediate and upper mass region. %due to the weaker explosion.
Only 25d-H shows a 
%$^{29}$Si-deficiency combined with the  $^{30}$Si-excess 
negative $\delta$($^{29}$Si/$^{28}$Si) combined with the positive $\delta$($^{30}$Si/$^{28}$Si)
in the deepest part of the He/C zone, as discussed in models 25T and 25av.
%SiC-X grains are highly biased to be $^{28}$Si-rich while LD graphites show part of the grains with $^{28}$Si-excess, but are mostly scattered around solar ratios. The condensation efficiency of graphite only depends on C enrichment, while that of SiC-X grains depends on both C and Si enrichment. Therefore, it is plausible that SiC-X grains will form more efficiently in the C/Si zone than in the He/C, while graphites only require a C-rich environment which applies to both the C/Si and the He/C zone.

\begin{figure*} %% C/Al plot %%%%%%%%%%%%%%%%%
    \centering
    \includegraphics[width=1.0\textwidth]{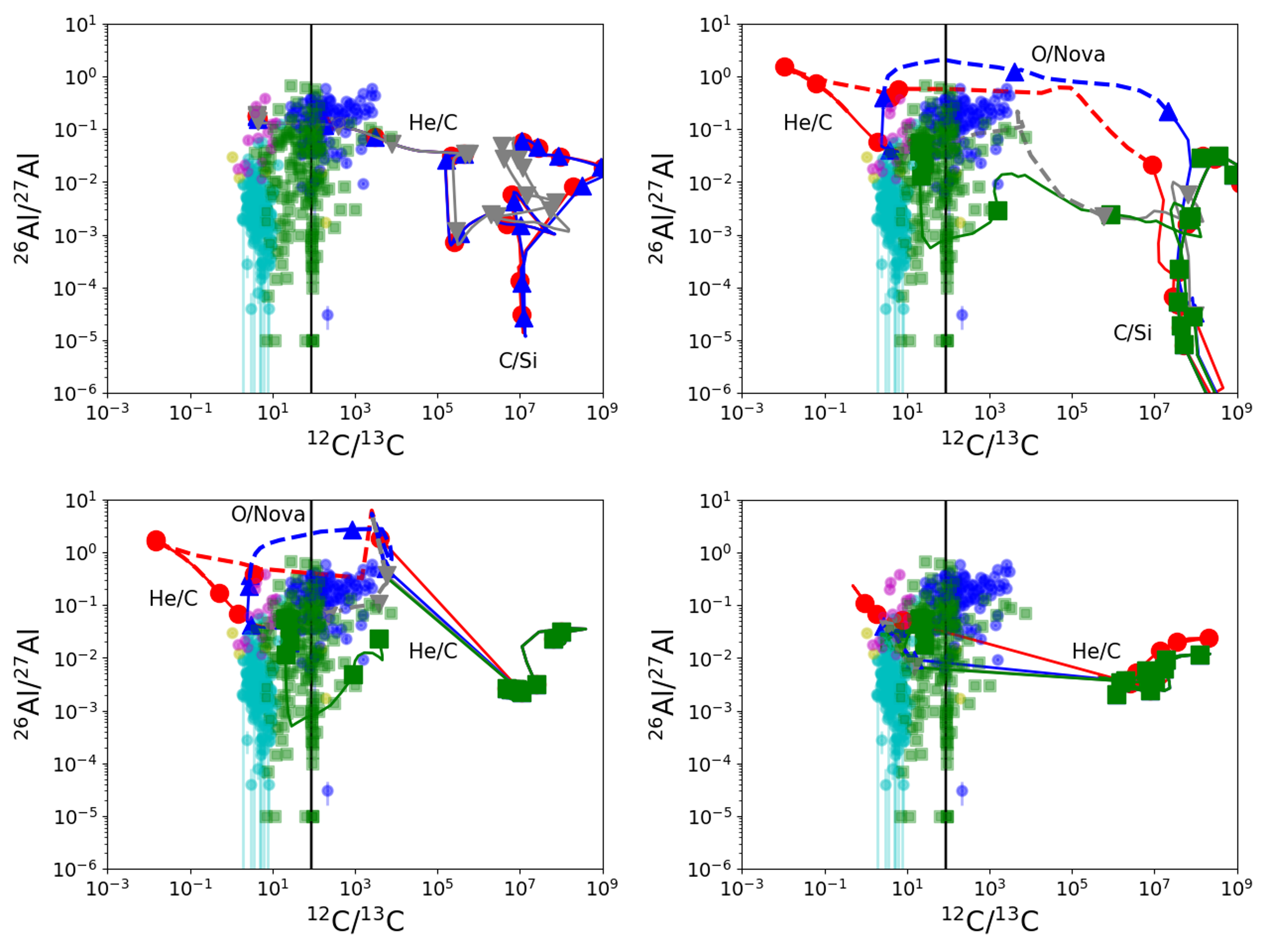}
    \caption{$^{26}$Al/$^{27}$Al and $^{12}$C/$^{13}$C isotopic ratios predicted by the CCSN models and compared with SiC grains (sub-types AB, nova, X, and C) and LD graphites are shown. For abundance ratios measured in grains and abundance profiles from stellar %models 
    simulations, symbols are used as in Figure \ref{fig:CvsN} in the four panels. Continuous lines represent the He/C and C/Si zones, while the dashed lines indicate the O/nova zone between them.}
    \label{fig:CvsAl}
\end{figure*}

Figure \ref{fig:CvsAl} presents the $^{12}$C/$^{13}$C and $^{26}$Al/$^{27}$Al isotopic ratios for %stellar models 
CCSN abundance simulations and grains. Compared to Figure \ref{fig:CvsN}, the full range of $^{12}$C/$^{13}$C ratio obtained in %stellar models 
the CCSN ejecta is shown. 
In %Panel 1
the top left Panel, models with different explosion energies but no H ingestion do not reach the $^{26}$Al enrichment needed to reproduce the full range of grains observations, as already discussed in \citet{Pignatari2013b} and \citet{Pignatari2015}. The comparison shown in %Panel 2 
the top right Panel has been already discussed by \citet{Pignatari2015} and we report them here for completeness. In particular, $^{12}$C/$^{13}$C shows a variation up to 11 orders of magnitude, with the highest ratios obtained in the C/Si zone on the right of the plot, to the lowest sub-solar ratios obtained at the intersection between the O/Nova zone and the He/C zone. The C-rich regions are plotted as solid lines whereas the O/nova zone is plotted as dashed lines. Based on preliminary calculations, we can anticipate that it is possible to reproduce the grains with highest $^{26}$Al/$^{27}$Al ratios through mixing of the He/C, O/nova, and C/Si zones of models 25T-H and 25T-H5. This is obtained even with the model 25T-H10 (not shown in the figure), considering mixtures of matter predominantly from the O/Nova zone with small contributions of matter from the C-rich zones below and above \citep[][submitted]{hoppe:22}. %, 25av-H, and 25av-H5. 

The bottom left Panel shows that the same high $^{26}$Al production during the CCSN explosion noted in %Panel 2 
the top right Panel can also be obtained in models at lower energy (25av), keeping the same H-ingestion parameterisation. In this case, the H left at the CCSN shock passage is controlling the $^{26}$Al production.
As with 25T-H5 compared to the other 25T models, 25av-H5 is the model showing the largest $^{26}$Al/$^{27}$Al ratio in the O/Nova zone, over the full range of measured C ratios. This is an interesting result. First, we would naively expect the $^{26}$Al enrichment due to explosive H-burning to increase with the amount of H left in the He shell. This is not the case for the 25T and the 25av stellar sets. Both $^{26}$Al and $^{27}$Al are made during the explosion in the O/Nova zone. Above a certain H abundance, $^{27}$Al is produced more efficiently than $^{26}$Al and the relative ratio decreases. In \cite{Pignatari2015}, we have seen that model 25T-H10 
(see Table \ref{tab:models2}) %(not included in this work) 
shows similar $^{26}$Al/$^{27}$Al ratios in the O/Nova zone compared to 25T-H. A maximum value of $^{26}$Al/$^{27}$Al is reached with H left between about 1\% and 1 per mill. More models would be needed to identify what is the amount of H yielding the largest $^{26}$Al/$^{27}$Al ratio, that could be higher than the abundances of 25T-H5. With the present models, the same conclusions can be derived for the 25av set at lower SN energies. By taking into account the measured scatter for C, N, Al and Si abundances in presolar grains (see also Figures \ref{fig:CvsN}, \ref{fig:dSi} and \ref{fig:dSi_zoom}), models 25T-H5 and 25av-H5 seem to provide the required conditions to match the typical H-burning signatures identified, and higher H concentrations do not help or make even worse the comparison with observations.

%As already seen in \citet{Pignatari2013b} and \citet{Pignatari2015}, the models with no H-ingestion do not reach the $^{26}$Al enrichment needed to reproduce the full range of grains observations.  
Finally, in %Panel 4 
the bottom right Panel the CCSN abundances ejected from model 25d cannot explain the highest $^{26}$Al/$^{27}$Al ratios observed. The temperatures reached in the He shell in these models during the SN shock passage are not high enough to efficiently make $^{26}$Al. The pre-SN $^{26}$Al abundance made during the H ingestion in our models is more than an order of magnitude too low compared to the presolar grains measurements of $^{26}$Al-rich SiC-X and graphites. %Results are really different between the 25av and 25d models. 
Beyond the scope of this paper, it would be useful to produce in the future one or two additional sets of CCSN simulations where the impact of the SN explosion energy is explored between 25d and 25av settings (see \S~\ref{sec:models}).

\begin{figure*} %% Ti44/dSi plot
    \centering
    \includegraphics[width=1.0\textwidth]{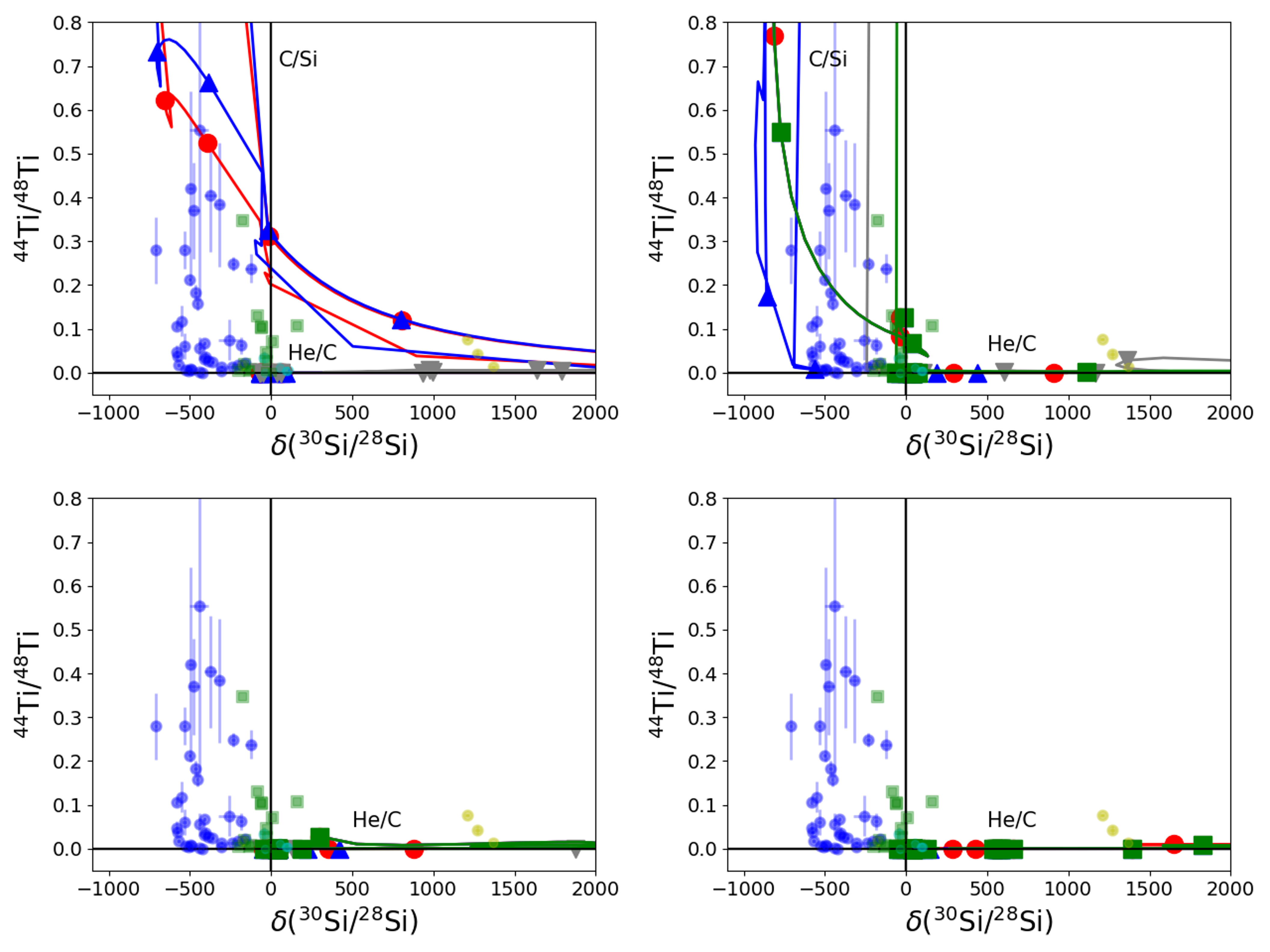}
    \caption{$^{44}$Ti/$^{48}$Ti and $^{30}$Si/$^{28}$Si isotopic ratios of SiC grains (sub-types AB, nova, X, and C) and LD graphites are compared with CCSN model predictions across the C-rich zones. For abundance ratios measured in grains and abundance profiles from stellar models symbols are used as in Figure \ref{fig:CvsN} in the four panels. The $\delta$-notation is used for the Si ratio.}
    \label{fig:Ti44vsSi}
\end{figure*}

Figure \ref{fig:Ti44vsSi} shows the $^{44}$Ti/$^{48}$Ti and $\delta$($^{30}$Si/$^{28}$Si) isotopic ratios for presolar grains and the four model sets. It is worth taking into consideration here that $^{48}$Ti and $^{48}$Ca are indistinguishable when measuring isotopic abundances in grains. However, $^{48}$Ca condenses much less efficiently into SiC than $^{48}$Ti \citep{Lodders1995}. 
The fractionation used for $^{48}$Ca/$^{48}$Ti is 0.0001, i.e. only 1 Ca atom condenses for every 10000 Ti atoms. In the Figure, therefore, the abundance of $^{48}$Ti also takes into account the contribution of $^{48}$Ca after fractionation.
The same fractionation is applied for $^{44}$Ca and the radioactive isotope $^{44}$Ti. As shown for 15d, 15r2, and 25T models in %Panels 1 and 2
the top Panels, $^{28}$Si-rich material coincides with high $^{44}$Ti %at high energies 
in the C/Si zone as they are both products of the explosive $\alpha$-capture chain \citep{Pignatari2013b}. In the central layers of the C/Si zone these models achieve high values of $^{44}$Ti/$^{48}$Ti, above the plotted area: the maximum ratios obtained are 3.78 and 2.97 for 15d and 15r2 respectively, and 29.9, 26.2, 26.1, and 26.1 for 25T-H, 25T-H5, 25T-H20, and 25T-H500 respectively (beyond the plot range). In particular for the 25T models, a large range of $^{44}$Ti/$^{48}$Ti is obtained in the C/Si zone, from no $^{44}$Ti to the high values mentioned above. As discussed in \citet{Pignatari2013b}, this variation of $^{44}$Ti production is a natural outcome of the $\alpha$-capture chain of reactions triggered by the explosive He-burning, where $^{44}$Ti is at the end of the production chain and requires high temperatures and $\alpha$-capture efficiency to be made. The milder $^{44}$Ti/$^{48}$Ti ratios observed in some $^{28}$Si-deficient graphites grains and SiC-C grains can be also reproduced in %Panel 1
the top left Panel, but not in the top right. %Panel 2.
%Notice also that the reaction path is not affected by the initial metallicity of the progenitor \citep{Pignatari2013c}. 
Instead, 15r4 %(Panel 1)
(top left Panel), 25av %(Panel 3)
(bottom left Panel), and 25d models %(Panel 4)
(bottom right Panel) are not energetic enough to form the C/Si zone with $^{28}$Si and $^{44}$Ti during the CCSN explosion. All of the models shown develop extremely high $\delta$($^{30}$Si/$^{28}$Si) in the He/C as the curve continues far past the right edge of the plot.

\begin{figure*}  %% Ti49/48 vs 46/48 %%%%%%%%%%
    \centering
    \includegraphics[width=1.0\textwidth]{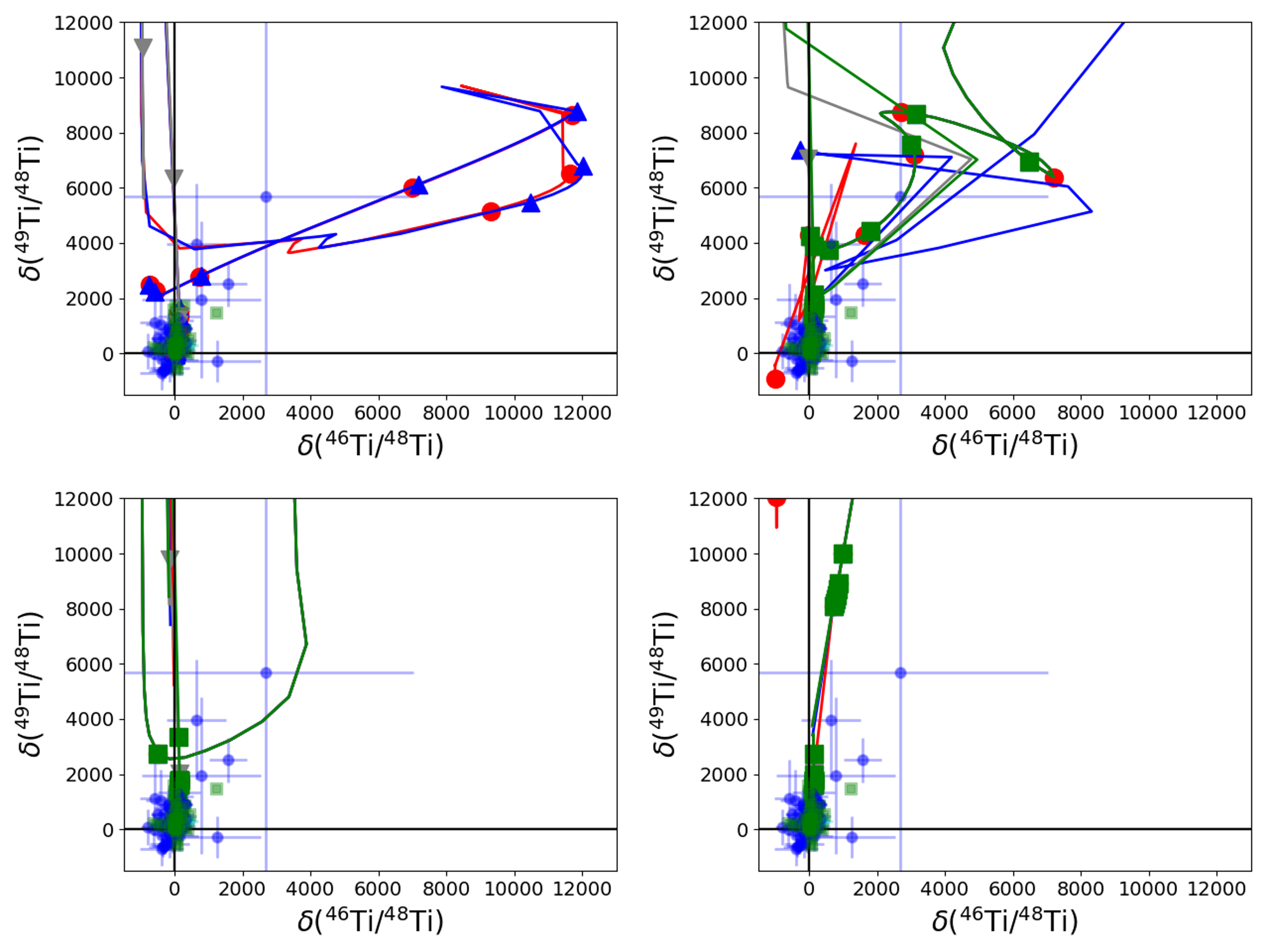}
    \caption{$^{49}$Ti/$^{48}$Ti and $^{46}$Ti/$^{48}$Ti isotopic ratios in $\delta$-notation of individual SiC grains (sub-types AB, nova, X, and C) and LD grains compared with abundance predictions of CCSN models. For abundance ratios measured in grains and abundance profiles from stellar models symbols are used as in Figure \ref{fig:CvsN} in the four panels.}
    \label{fig:Ti49}
\end{figure*}

Figure \ref{fig:Ti49} shows the $^{49}$Ti/$^{48}$Ti and $^{46}$Ti/$^{48}$Ti ratios for our models compared to presolar grains. The isotope $^{49}$V is a radioactive parent of $^{49}$Ti, with a half-life of 330 days. This means that some $^{49}$V remains when the grains are condensing. Vanadium (in the form of VC$_{0.88}$) and Ti (as TiC) are incorporated in ideal solid solution into SiC at the same 50\% condensation temperature \citep{Lodders1995}. Therefore, $^{49}$V can be expected to condense at a similar rate to $^{49}$Ti in SiC grains and graphites, and we do not apply any fractionation between V and Ti in Figure \ref{fig:Ti49}. 
Instead, the contributions from the stable isobars $^{46}$Ca and $^{48}$Ca to $^{46}$Ti and $^{48}$Ti respectively, are taken into account as in Figure \ref{fig:Ti44vsSi}. %the fractionation this work uses for $^{49}$V/$^{49}$Ti is 1, i.e. 1 $^{49}$V atom condenses for every $^{49}$Ti. 
All of the CCSN stellar sets presented in Figure \ref{fig:Ti49} show the capability to produce high $^{49}$Ti/$^{48}$Ti ratios in the C-rich ejecta. In %Panel 1
the top left Panel, for the C/Si zone 15d and 15r2 make $^{49}$Ti mostly as $^{49}$V. An additional $^{49}$Ti component is made in the He/C zone by neutron captures, directly as $^{49}$Ti or as radiogenic contribution from the neutron-rich radioactives $^{49}$Ca and $^{49}$Sc. Models in 
the top right Panel %Panel 2 
still have the $^{49}$V-rich component in the C/Si zone, but neutron captures in the He/C zone are suppressed in the models with most H left. However, in 25T-H and 25T-H5 the abundant pristine $^{48}$Ti is partially converted by direct proton captures to $^{49}$V in the O/Nova zone and at the bottom of the He/C zone during the CCSN explosion, generating also in these conditions a super-solar $^{49}$Ti/$^{48}$Ti ratio ($^{48}$Ti and $^{49}$Ti are respectively 73.72\% and 5.41\% of the solar Ti). The 25av models %(Panel 3) 
(bottom left Panel) and 25d models %(Panel 4) 
(bottom right Panel) also reproduce the observed scatter measured in presolar grains. Therefore, depending on the SN explosion energy and on the occurrence of H-ingestion, we can identify three components yielding super-solar $^{49}$Ti/$^{48}$Ti ratio: a $^{49}$V-rich one in the C/Si zone due to explosive He burning, a (possibly) $^{49}$V-rich one in the O/Nova zone (and in the deepest He/C zone layers) due to proton-captures and, if no significant amount of H is present at the onset of the SN explosion, the $^{49}$Ti-rich neutron-capture component in the He/C zone. The neutron-burst in the He/C zone is suppressed in one-dimensional CCSNe models developing the O/Nova zone \citep[][]{Pignatari2015}. Therefore, if we do not consider the mild pre-SN s-process signature in the He shell, the proton-capture and the neutron-capture components discussed above could be alternative results of local explosive nucleosynthesis. However, the multi-dimensional dynamic nature of H ingestion could easily lead to asymmetries and local differences in the He shell which could make the two components complementary products of the same CCSN explosion. While there are several studies of H ingestion with multi-dimensional hydrodynamics simulations for low-mass stars \citep[][]{stancliffe:11, herwig:14, woodward:15}, at present the same extensive studies are not available for massive stars to better constrain the stellar progenitors \citep[see e.g.,][]{clarkson:21}. The present limitations of one-dimensional models used in this work need to be taken into account, and therefore we cannot derive strong constraints about the relative relevance of these difference components at the moment. Based on the same generation of models, \cite{liu:18} used the $^{49}$Ti signature to distinguish between the contribution of the C/Si and the Si/S zones to condense SiC-X grains. They argued that by invoking the Si/C zone as the main source of $^{28}$Si, the $^{49}$Ti from the He/C zone would be the dominant contributor to the $^{49}$Ti measured in grains, at odds with the data. Their analysis was based on considering the Ti/Si ratio of the Si/C zone, which is orders of magnitude lower than that of the He/C zone in CCSN models. Given what we have said above and taking into account the uncertainties involved, we do not think these conclusions can be driven yet just based on the Ti/Si ratio. 
In particular, while the abundance of C in the Si/S region is negligible, the abundances of both C and Si are higher by at least an order of magnitude in the C/Si zone compared to the He/C zone. In the stellar progenitor the He/C zone is only a partial He-burning region (the former convective He shell), and at the onset of the collapse still a lot of He fuel is left: $^{4}$He is the most abundant isotope in the He/C zone. The C/Si zone is located instead at the bottom of the former He shell, where hydrostatic He-burning was almost complete. This is the reason why C can be even an order of magnitude higher than in the He/C zone. Additionally, in this region explosive He-burning is destroying the pre-SN O to make Si and other elements, while C is not depleted because the $^{12}$C($\alpha$,$\gamma$)$^{16}$O rate is orders of magnitudes weaker than the $^{16}$O($\alpha$,$\gamma$)$^{20}$Ne reaction rate \citep[][]{Pignatari2013b}. 
Therefore, independently from the Ti/Si ratio, the high enrichment in both C and Si is generating a qualitative expectation that ejecta from the C/Si zone will have a much higher formation efficiency of SiC grains compared to the He/C zone. Ti should then condense in these grains from consistent mixtures of C-rich and Si-rich stellar material. More detailed predictions for Ti chemistry in environments with different degrees of C and Si enrichment would require complete network calculations with chemical reactions, that are beyond the scope of this paper.
%All of the models, including the lower energy 25av and 25d, also show a region of high $^{49}$Ti likely produced by neutron-capture on lighter Ti isotopes.

For the sake of completeness, in the Appendix Figures \ref{fig:Ti47} and \ref{fig:Ti50} provide the isotopic ratios $^{47}$Ti/$^{48}$Ti with respect to $^{46}$Ti/$^{48}$Ti, and $^{50}$Ti/$^{48}$Ti with respect to $^{46}$Ti/$^{48}$Ti for our models in comparison to presolar grains. No additional constraints can be derived from these figures compared to what we have already discussed. Nevertheless, as for Si isotopes and other Ti isotopes discussed so far, a novel nucleosynthesis in the O/Nova zone compared to massive star models without H ingestion can be also identified. We defer to a future paper for the detailed analysis of the nucleosynthesis in the O/Nova zone.

\section{The abundance signatures of radioactive isotopes $^{26}$Al and $^{44}$Ti in presolar grains} %%%%%%%%%%%%%%%%%%%%%%%%%%%%%%%
\label{sec:discussion}

In this section we look further at the implications of the models predictions, and for each grain type we consider the distributions of isotopic abundance ratios for the radioactive isotopes $^{26}$Al and $^{44}$Ti. 

\begin{figure}  %% stats C vs Al
    \centering
    \includegraphics[width=0.47\textwidth]{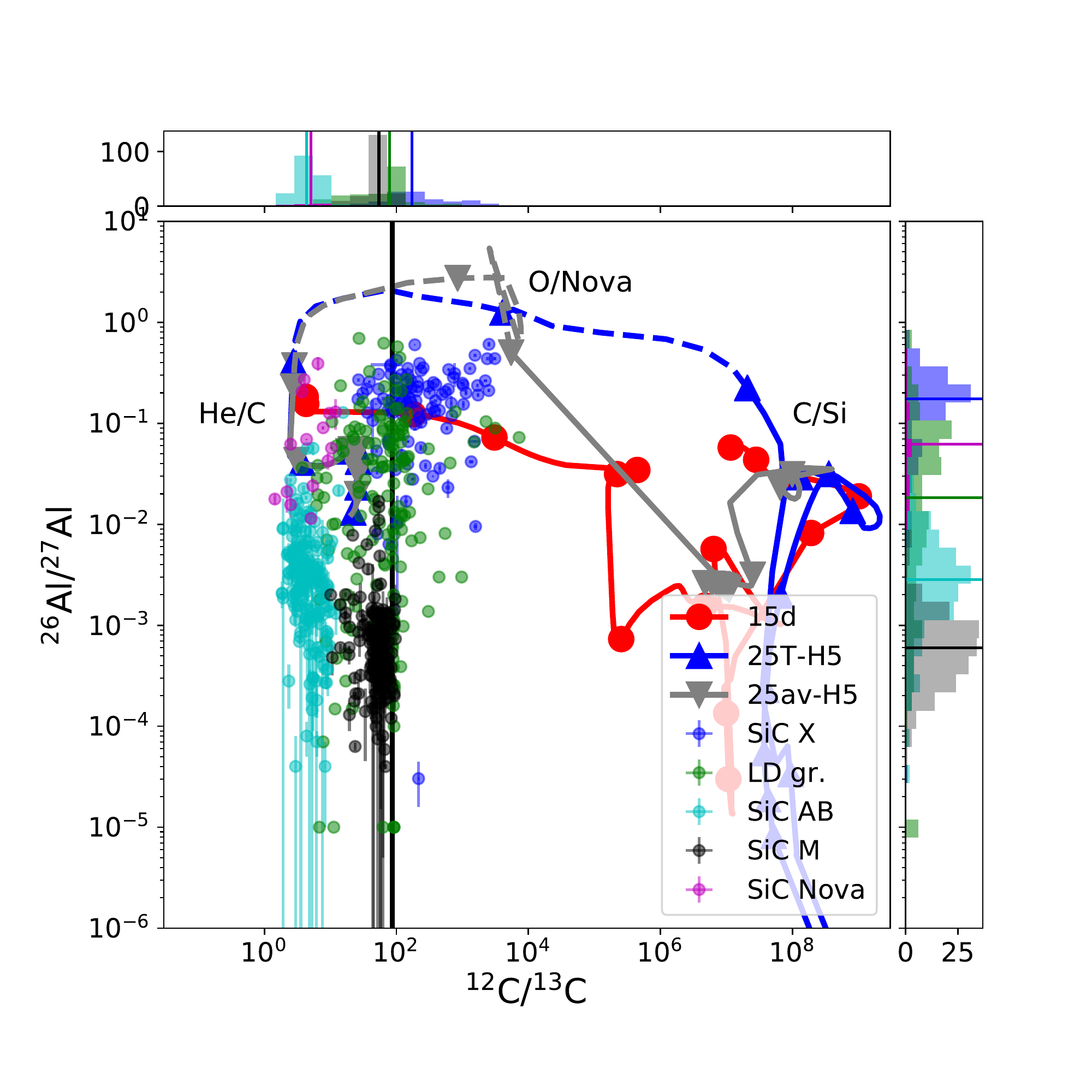}
    \includegraphics[width=0.47\textwidth]{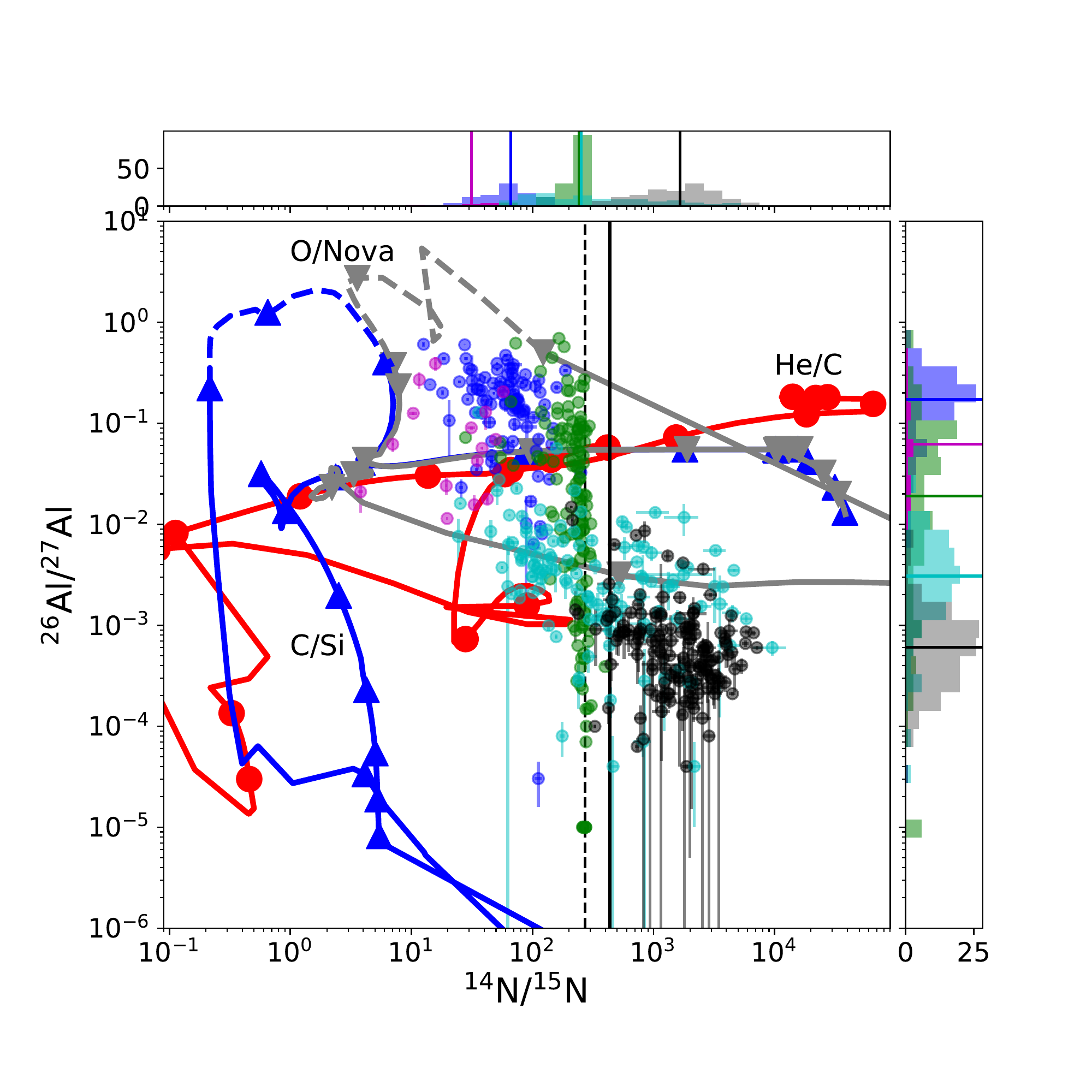}
    \caption{%Panel 1
    Top Panel: $^{26}$Al/$^{27}$Al and $^{12}$C/$^{13}$C isotopic ratios predicted from 15d, 25T-H5 and 25av-H5 are compared with SiC grains (178 sub-types AB, 15 nova, and 106 X) and 173 LD graphites. The presolar grains data are available from \citep{Hynes2009,Stephan2020}. As a reference, we also show the data for 185 SiC mainstream grains (labelled as M), which are the most abundant type of SiC presolar grains and they are made in Asymptotic Giant Branch stars \citep[][and references therein]{Zinner2014}. %For each grain type, the number of grains (n) and the median $^{26}$Al/$^{27}$Al (m) are shown. 
    The median $^{26}$Al/$^{27}$Al ratios are 0.1740, 0.0028, 0.0620 and 0.0006 for SiC-X, AB, Nova and Mainstream, respectively. The median value for LD graphites is 0.0184.
    Continuous lines on the main plot represent the He/C and C/Si zones, while the dashed lines indicate the O/nova zone between them. The histograms on the axes show the distribution of each grain type in their respective isotopic ratio, with a solid opaque line indicating the median ratio. %Panel 2
    Bottom Panel: the same as in %Panel 1
    the top panel, but for $^{26}$Al/$^{27}$Al and $^{14}$N/$^{15}$N isotopic ratios. Both solar and terrestrial N ratios are shown, as in Figure \ref{fig:CvsN}.}
    \label{fig:stat_C_and_NvsAl}
\end{figure}

Figure \ref{fig:stat_C_and_NvsAl}, %Panel 1
top Panel, shows the $^{26}$Al/$^{27}$Al and $^{12}$C/$^{13}$C isotopic ratios predicted by a selection of our CCSN models 15d, 25T-H5 and 25av-H5 and compared with presolar grains.
The distribution of isotopic ratios for each grain type is shown and the respective median values are reported. Based on the considerations made in the previous section, we used here for the analysis 25T-H5 and 25av-H5, and not the analogous CCSN models with the highest initial H concentration. Note however that the results that we derive from the discussion would not be much different. 
As we have seen in Figure \ref{fig:CvsAl}, the ejected abundance products from 25T-H5 and 25av-H5 have a sub-solar $^{12}$C/$^{13}$C ratio in the He/C zone above the O/Nova zone. In both of the two models, the ratio is becoming super-solar moving inward within the O/Nova zone, and with a depletion of both C isotopes by proton captures. At the bottom of the O/Nova zone, both the C/Si zone in 25T-H5 and residual C-rich layers in 25av-H5 are $^{12}$C-rich. In the 15d model, both the C/Si zone and the He/C zone have a $^{12}$C/$^{13}$C ratio higher than solar, while the upper part of the He/C zone and the He/N zone are sub-solar \citep{Pignatari2013b}.

If now we check more carefully the abundance distributions in single presolar grains, SiC-X grains appear to show two $^{26}$Al enrichment peaks, one distinct at around $^{26}$Al/$^{27}$Al $\approx$ 0.2 and potentially a second peak (less clear with fewer data points) at $^{26}$Al/$^{27}$Al $\approx$ 0.05. More data points would be required to clarify whether these are indeed two distinct peaks or simply a single distribution with a long tail toward low $^{26}$Al abundances. Based on \cite{Pignatari2015} and results presented in \S~\ref{sec:results}, the first peak may be the signature of late H-ingestion and SN explosions at higher energy (25T or 25av). The potential second peak could be explained from mixing of the He/C and He/N zones (15d reaches these values), or H-ingestion in low-energy CCSNe (e.g. 25d, see Figure \ref{fig:CvsAl}). This should be confirmed by future works, where mixing models are applied to reproduce the composition of single presolar grains. The first peak is much stronger, which would be consistent with the scenario where most of SiC-X are forming in high-energy explosions \citep[][]{Pignatari2013b}. We cannot derive a direct constraint on the frequency of H-ingestion in CCSNe. This is because most of the observed CCSN dust could have originated from a single CCSN immediately before the formation of the Solar System, with late H-ingestion and asymmetric ejecta. The distribution of $^{26}$Al/$^{27}$Al measured in LD graphites is quite different to that of SiC-X grains, despite the overall similar range of ratios found. This may appear to be quite surprising, considering that the two types of grains are formed in the same parts of the CCSN ejecta. There is a higher peak at $^{26}$Al/$^{27}$Al $\approx$ 0.06 and a shallow broad peak at $\approx$ 0.01. 
Like for SiC-X grains, it is still unclear if this second peak really exists. However, the distribution of $^{26}$Al/$^{27}$Al ratios appear to be shifted toward higher values for SiC-X compared to LD graphites. This could be explained assuming that LD graphites condensed efficiently in CCSN ejecta exposed to a broader range of nucleosynthesis conditions compared to SiC-X grains, including less extreme temperature and density peaks during the SN explosion \citep[e.g.,][]{Pignatari2018}. On the other hand, as \cite{Travaglio1999} pointed out we cannot exclude that the $^{26}$Al/$^{27}$Al measured today in graphites was affected by loss of radiogenic $^{26}$Mg, which would also explain these lower ratios.
%Both SiC and graphite require C/O > 1 for condensation, so I find it hard to explain the higher 26Al/27Al ratios of SiC X grains compared to LD graphites. SiC formation requires significant amounts of Si, e.g., from the Si/S zone, but there 26Al/27Al is comparatively low which makes things even worse. Travaglio et al. 1999 considered  loss of radiogenic 26Mg from the graphites; maybe this is still the best explanation.
SiC-AB grains show a broad distribution with a peak at 0.003. Since different types of stellar sources are contributing to these grains \citep[][]{Amari2001,Liu2017,Liu2017b,Hoppe2019}, we cannot derive clear constraints applying specifically to CCSNe from the overall distribution shown.
Possibly because of the low number of data, the  $^{26}$Al/$^{27}$Al in putative Nova grains is scattered over two orders of magnitudes, partially overlapping with the SiC AB grains, and with a median value of 0.062 that is consistent with the LD graphites higher peak. 

In %Panel 1
the top Panel, we also report the abundance distribution with the position of the median line for the measured $^{12}$C/$^{13}$C ratios. Interestingly, as we have seen for Al isotopes, we find again that the C distribution for SiC-X and LD graphites is different, with the median at 172 (higher than solar) and 79 (lower than solar) respectively. This is an intrinsic difference of the two populations of grains, and it is clearly not due to a different solar contamination or loss of radiogenic material as it could be for $^{26}$Mg. In the figure, we report 106 SiC-X grains with 22\% of $^{12}$C/$^{13}$C ratios less than solar and 78\% $^{12}$C/$^{13}$C greater than solar, and 173 LD graphites with 52\% of $^{12}$C/$^{13}$C ratios less than solar and 48\% $^{12}$C/$^{13}$C greater than solar. In the classic scenario where $^{28}$Si is formed in the  Si/S zone, there is no explanation of these differences, since the Si/S zone is C-poor (and Al-poor) and would not affect the C-ratio. Additionally, if we assume that both $^{13}$C and $^{26}$Al found in grains are a product of explosive H-burning or of the He/N zone, because the $^{12}$C/$^{13}$C median of LD graphites is lower than in SiC-X data we would also expect the $^{26}$Al/$^{27}$Al median to be higher. This is not the case.  On the other hand, this appears to be further confirmation of the C/Si zone as the origin of $^{28}$Si in these grains. Compared to LD graphites, a larger fraction of SiC-X are $^{28}$Si-rich and $^{12}$C-rich like the C/Si zone. This could qualitatively explain why a larger fraction of SiC-X grains have $^{12}$C/$^{13}$C larger than solar, independently from the nucleosynthesis source of $^{13}$C and $^{26}$Al. The SiC-X grains preferentially form in high-energy CCSNe or in high energy components of asymmetric CCSNe ejecta where the C/Si zone is created during the explosion. On the other hand, LD graphites are formed in a larger range of conditions, in ejecta with and without the $^{28}$Si-rich and $^{12}$C-rich C/Si zone. Finally, SiC nova grains and SiC-AB grains show consistent $^{12}$C/$^{13}$C distributions, with medians equal to 5.1 and 4.4, respectively.

In figure \ref{fig:stat_C_and_NvsAl}, %Panel 2
bottom Panel, we show the $^{26}$Al/$^{27}$Al and $^{14}$N/$^{15}$N isotopic ratios from the CCSN abundances predicted in 15d, 25T-H5 and 25av-H5 and we compare them with SiC grains. Models 25T-H5 and 25av-H5 show the same abundances in the external part of the He/C zone, with an $^{26}$Al/$^{27}$Al of the order of 0.05, generated from the H-ingestion event, and a $^{14}$N/$^{15}$N ratio higher than solar, but quickly decreasing with moving deeper in the He/C zone.  
At the bottom of the He/C zone a sudden rise of the $^{26}$Al/$^{27}$Al up to about a value of 0.3 is also seen for both the two models. We can clearly see that in these regions nucleosynthesis patterns are shaped by nuclear reactions, which are the same in the two models. The $^{26}$Al/$^{27}$Al ratio rises up to 2-3 in the O/Nova zones of both 25T-H5 and 25av-H5. However, the N isotopic ratio decreases in 25T-H5 and increases in 25av-H5. For the presolar grains analysis of the N abundances the bottom of the O/Nova zone and the C/Si zones are not relevant, since they are both N-poor. 
The model 15d shows higher ratios in the external He/C zone compared to the two models with H-ingestion, reaching a maximum $^{26}$Al/$^{27}$Al value of 0.2 but with much lower N abundances \citep[][]{Pignatari2015}. In the He/C zone and in the C/Si zone $^{26}$Al is not made efficiently by explosive nucleosynthesis. The explosive production of $^{15}$N allows the reduction of the N isotopic ratio, but its production is too weak compared to observations \citep[][]{bojazi:14}, and the whole N abundances in the deepest C-rich ejecta shown in the figure would not be relevant once some small degree of mixing is allowed for ejected stellar material.

We have already discussed in %Panel 1
the top Panel the $^{26}$Al/$^{27}$Al distribution in presolar grains. The $^{14}$N/$^{15}$N for LD graphites is not indicative of the original composition because of terrestrial contamination. Where contamination from terrestrial material did not completely erase the N signature, the LD graphites carry a low N ratio, particularly for high $^{26}$Al/$^{27}$Al ratios. The original $^{15}$N-rich signature was likely very high in these grains. 
%Without a detailed grain-by-grain study, it cannot be known whether this a nucleosynthesis signature or the effect of more limited mixing before grain condensation.
Concerning the different types of SiC grains shown in the figure, there is a general trend (due to nuclear astrophysical reasons) of correlated $^{15}$N and $^{26}$Al abundances across the whole grain population, but within a large grain-to-grain scatter. Putative nova grains again behave like a $^{15}$N-rich tail of the SiC-AB grain spread ($^{14}$N/$^{15}$N medians of SiC Nova and SiC-AB grains are 31 and 252, respectively). The population of SiC-X grains carry a higher $^{26}$Al compared to other grains with the same N ratio. The sub-solar $^{14}$N/$^{15}$N ratios show a scatter of a factor of 20, with a median of 66 that is in between SiC Nova and SiC-AB grains.

%\begin{figure}  %% stats N vs Al
%    \centering
%    \includegraphics[width=0.47\textwidth]{figures/test_stat_n_vs_al26al27.pdf}
%    \caption{$^{26}$Al/$^{27}$Al and $^{14}$N/$^{15}$N isotopic ratios predicted by the 25T-H and 15d models and compared with SiC grains (sub-types mainstream, AB, nova, and X) and LD graphites. For each grain type, the number of grains with both ratios measured (n) and the mean $^{26}$Al/$^{27}$Al of those grains (m) are shown. Continuous lines on the main plot represent the He/C and C/Si zones, while the dashed lines indicate the O/nova zone between them. The histograms on the axes show the distribution of each grain type in their respective isotopic ratio, with a solid opaque line indicating the mean ratio.}
%    \label{fig:stat_NvsAl}
%\end{figure}

\begin{figure}  %% stats Ti44 vs Al
    \centering
    \includegraphics[width=0.47\textwidth]{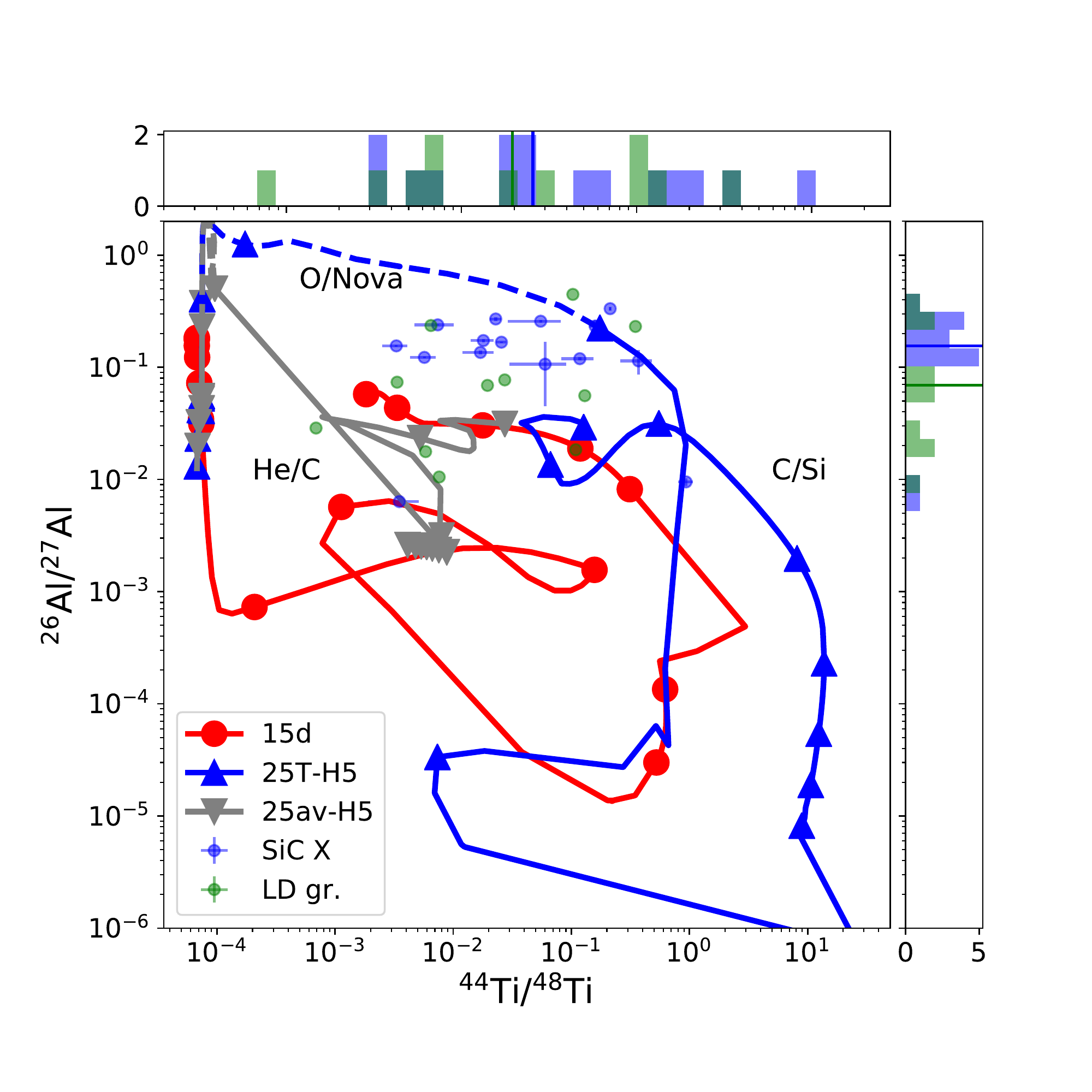}
    \caption{$^{26}$Al/$^{27}$Al and $^{44}$Ti/$^{48}$Ti isotopic ratios predicted from the CCSN models 15d, 25T-H5 and 25av-H5 are compared with 15 SiC-X grains and 11 LD graphites. The median $^{26}$Al/$^{27}$Al ratios are 0.16 and 0.07, respectively. Continuous lines on the main plot represent the He/C zone (squeezed at $^{44}$Ti/$^{48}$Ti $\lesssim$ 10$^{-4}$) and the C/Si zone, while the dashed lines indicate the O/nova zone between them. The histogram on the $^{26}$Al/$^{27}$Al axis shows the distribution of each grain type, with a solid opaque line indicating the mean ratio. Also a histogram of $^{44}$Ti/$^{48}$Ti ratios is shown.}
    \label{fig:stat_Ti44vsAl}
\end{figure}

Figure \ref{fig:stat_Ti44vsAl} shows the $^{26}$Al/$^{27}$Al and $^{44}$Ti/$^{48}$Ti isotopic ratios from the final CCSN abundances in 15d, 25T-H5 and 25av-H5 and compared with SiC-X and LD graphites.
%by the the 25T-H, 25T-H5 and 15d models and compared with SiC-X grains and LD graphites. 
%Since there are much fewer SiC-X grains and LD graphites here (15 and 11 respectively), we will not consider the histogram aspect of this figure. 
Only a limited number of grains have been measured for both $^{26}$Mg ($^{26}$Al) and $^{44}$Ca ($^{44}$Ti): 15 SiC-X and 11 LD graphites. A particular problem for Ca isotope measurements is contamination of the grains with terrestrial Ca, such that existing $^{44}$Ca excesses may remain hidden \citep[][]{Besmehn2003}. This may explain that from the 138 X grains measured for Ca, only 38 showed resolvable excesses in $^{44}$Ca, although it should be noted that Ca contamination has no effect on inferred $^{44}$Ti/$^{48}$Ti ratios. Because of the low statistics available, we cannot derive any significant conclusions from the abundance distribution in the histogram aspect of this figure. It is interesting to note that for SiC-X grains the $^{26}$Al/$^{27}$Al medians between this sample and the extended sample from Figure \ref{fig:stat_C_and_NvsAl} are consistent within 12\%. For LD graphites, the 11 grains here have a median that is a factor of 3.7 higher compared to the extended sample in Figure \ref{fig:stat_C_and_NvsAl}. However, we cannot yet conclude that $^{44}$Ti-rich LD graphites tend to be also $^{26}$Al-rich based on 11 grains only. More grains with both $^{44}$Ti and $^{26}$Al measurements would be needed.

Concerning the abundance curves from CCSN models shown in Figure \ref{fig:stat_Ti44vsAl}, the C/Si zone in 15d and 25T-H5 provides the $^{44}$Ti, with $^{44}$Ti/$^{48}$Ti ratios larger than 0.001 \citep[see e.g.,][]{Pignatari2013b}.  The C/Si zone does not make $^{26}$Al. Instead, in the figure this radioactive isotope is provided by the O/nova zone (in 25T-H5 and 25av-H5) and at the bottom of the He/C zone (the region around $^{44}$Ti/$^{48}$Ti $\lesssim$ 0.0001). We may anticipate here that a local mixing of these regions would result in a large degree of $^{26}$Al and $^{44}$Ti enrichment. In the future, extensive comparisons between mixing models from CCSN models with H ingestion and single presolar grains carrying both $^{44}$Ti- and $^{26}$Al-enrichment are needed. Previously, \cite{xu:15} used the models 15r (really similar to the model 15d presented here) and 15r4 (presented in this work) to reproduce a sample of single SiC X grains including both $^{26}$Al and $^{44}$Ti. However, both these models are without H ingestion. In \cite{Hoppe2018}, for the first time $^{44}$Ti was explored for two SiC-X grains and a SiC-C grain in the context of the 25T H-ingestion SN models presented here. From that study, it was shown that predictions for $^{44}$Ti are too high when C-, N-, and Si-isotopic compositions and $^{26}$Al/$^{27}$Al ratios are well fitted. The possible sources of such a discrepancy were considered, among others the extension and detailed abundance profiles of the C/Si zone.  
%We cannot see $^{26}$Al enrichment above the O/nova zone as a nuclear astrophysical boundary is hit.
Interestingly, in the high energy CCSN model 25T-H5 we also see a $^{44}$Ti production in the O/nova region, with $^{44}$Ti/$^{48}$Ti increasing up to about 0.2 with decreasing mass coordinate. The absolute abundance of $^{44}$Ti in this region is about three orders of magnitudes lower than in the C/Si zone, reaching a maximum abundance in the order of 10$^{-6}$ in mass fraction. The nucleosynthesis in the deepest layers of the O/Nova zone is given by the combination of proton captures and $\alpha$-captures, that start to become relevant at these conditions. Some $^{44}$Ti is made and at the same time the initial $^{48}$Ti is depleted by a factor of 2-3, which explains the rise of the $^{44}$Ti/$^{48}$Ti ratio.%  as well as the C/Si zone. 
%In the O/nova region, there is also a depletion of $^{42}$Ca and a production of $^{43}$Sc, so this is likely a proton-capture signature of light Ca isotopes. While the absolute abundance of $^{44}$Ti is not as high here as in the C/Si zone, $^{48}$Ti has approximately solar abundance and $^{44}$Ti is produced enough to cover the range of ratios observed in the grains. % very rough trend with Ti44/48Ti ratios decreasing with increasing Ti content. (see Peter email from 3 Nov 2020)

%Due to the relatively high abundance of both Ti and C there, the C/Si zone is a good candidate for the site of TiC grain condensation. However, this is beyond the scope of this study.

\section{Conclusions} %%%%%%%%%%%%%%%%%%%%%%%%%%%%%%%%%%%%%%%%%%%%%%%%%%%%%%%%%%%%%

This work has compared the isotopic abundances of presolar SiC grains and graphites from CCSNe with the predictions of four sets of one-dimensional CCSN models with a range of energies, both with and without H-ingestion: 15d, 25T, 25av and 25d. Compared to previous works, we are including in our analysis six additional CCSN models with H ingestion (set 25av), and intermediate explosion energies between the 25T and 25d stellar sets \citep[][]{Pignatari2015}. The primary aim of this work is to study the impact of differing energies and remaining H ($0.0024\% \leq X_H \leq 1.2\%$) in H-ingestion models as well as comparing the predictions of models with and without H-ingestion but with the same range of explosion energies.
The isotopic ratios focused on were C, N, Al, Si, and Ti of LD graphites and SiC grains of Type X, AB, C and nova, which have potentially condensed from CCSN ejecta. In particular, we discuss the production of the radioactive isotopes $^{26}$Al and $^{44}$Ti.

The 15d, 15r2, and 25T models develop a C/Si zone, ranging in explosion energy between 1-$5 \times 10^{51}$erg and with a temperature at the bottom of the C-rich region between $2.2 \times 10^9$K (15r2) and $2.3 \times 10^9$K \citep[25T;][]{Pignatari2015}. These models are able to reproduce the $^{28}$Si- and $^{44}$Ti-excesses of SiC-X grains and LD graphites, consistently with the nucleosynthesis signatures of Si and Ti isotopes measured. The most H-rich 25av models still develop an O/Nova zone, but the C-rich layers ejected at the bottom of the region are not forming a fully built C/Si zone: the peak temperature of $1.5 \times 10^9$K was not high enough to efficiently convert $^{16}$O into $^{28}$Si. However, we do see the destruction of $^{16}$O here as well as a small signatures for $^{20}$Ne and $^{24}$Mg production.
The nucleosynthesis signatures calculated in the O/Nova zone shows remarkable similarities in the 25T and 25av sets, showing that in principle it is possible to disentangle the impact of the amount of H present in the He shell at the CCSN onset from the impact of the CCSN explosion energy. The main reason is that by increasing the explosion energy the O/Nova zone is extending more outward in the former convective He shell, mitigating the effect of the higher temperatures generated. For both the two CCSN sets 25T and 25av, we find that the models with initial $X_H \approx 0.0024$ (25T-H5 and 25av-H5) are producing the highest $^{26}$Al/$^{27}$Al ratios, reproducing the most extreme $^{26}$Al enrichment measured in grains. In particular, the $^{26}$Al/$^{27}$Al ratio is not increasing linearly with the increasing of the initial H, independently from the explosion energy. We show that both the two sets of models develop Si isotopic ratios at the bottom of the He/C zone that are consistent with SiC-Nova grains, together with the C and N ratios (models 25T-H, 25T-H5 and 25T-H10, 25av-H, 25av-H5 and 25av-H10). 
On the other hand, the nucleosynthesis of Si isotopes in the O/Nova zone is quite different between 25T-H (25av-H) and 25T-H5 (25av-H5), with the highest H abundances driving an efficient production of both $^{29}$Si and $^{30}$Si. We discuss some of the relevant similarities between the nucleosynthesis obtained in the O/Nova environment compared to ONe Novae \citep[e.g.,][]{amari:01}. 

Finally, we discussed the abundance distribution in presolar grains, looking in particular at C and N isotopic ratios, $^{26}$Al and $^{44}$Ti. Two distinct populations can be possibly identified for both SiC-X grains and LD graphites, an $^{26}$Al-rich one and an $^{26}$Al-poor. However, the $^{26}$Al-poor component is less clearly defined, and more statistics is needed. in order to draw meaningful conclusions.
The C abundance distribution instead show relevant differences between SiC-X grains and LD graphites. Combined with the Al signature, we discuss and found these distributions consistent with the scenario where the $^{28}$Si-rich contribution would come from the $^{12}$C-rich C/Si zone. On the other hand, the same explanation would not apply if the $^{28}$Si-rich contribution would come from the $^{12}$C-poor Si/S zone. Unfortunately, there is only a limited number of grains with both $^{26}$Al and $^{44}$Ti measured. More data would be required to derive trends between $^{26}$Al and $^{44}$Ti enrichments.

%Specifically, 25T-H and 25T-H5 were able to reproduce the most $^{28}$Si-rich, $^{44}$Ti-rich grains for the first time, suggesting that a C/Si zone is needed to form them. While not being energetic enough to reproduce some of the signatures (e.g. $^{28}$Si, $^{44}$Ti), the new 25av models 25av-H and 25av-H5 were able to reproduce the highest $^{26}$Al/$^{27}$Al ratios seen in these grains.

In future works, we plan to study in detail the nucleosynthesis in the O/Nova zone for the models discussed here. We identify significant differences in the production of Si and Ti isotopes by reducing from initial H concentrations in the order of 1\% (models 25T-H and 25av-H) to few per mill (models 25T-H5 and 25av-H5). While the production path of the radioactive isotope $^{26}$Al is similar, the resulting $^{26}$Al/$^{27}$Al ratio does not behave linearly with the initial abundance of H explored.  Additionally, these models need to be further tested by comparing their predictions with elements and isotopic ratios on a grain by grain basis, where the qualitative consistency of the abundance signatures described in this work need to be quantitatively compared within mixtures of CCSN layers to match the measured data \citep[e.g.,][]{Travaglio1999,xu:15, Liu2016, Liu2017, Liu2018, Hoppe2018, Hoppe2019, hoppe:21}. This would act as further diagnosis of these models and allow the continued improvement upon them where grain signatures are not reproduced. The network of nuclear reactions used for our stellar calculations is becoming more reliable at the temperatures range relevant for explosive He-burning \citep[see however, e.g.,][]{bojazi:14,Pignatari2013c}. The CCSN explosion energy as a single parameter to explore from the different nucleosynthesis patterns is relatively easy to handle by specific nuclear astrophysics studies. However, as we stressed in \cite{Pignatari2015} and other more recent works the outcome of H ingestion events in one-dimensional models should be taken only as a qualitative guidance in specific comparisons with single presolar grains. While proton-capture rates are typically well known at temperatures in the order of 2-3$\times$10$^8$ K, multi-dimensional hydrodynamics models are needed to define the evolution of the structure of He-burning layers, the local nucleosynthesis driven during the H ingestion and eventually the amount of H left in the He-burning ashes. To this end, presolar grains represent the most powerful observational diagnostic available to study these events in massive star progenitors of past CCSNe.
%Comparing the models with heavy nuclei such as Mo, Ru, and Ba would give an indication of their performance in n-process calculations.

\section{Acknowledgements} %%%%%%%%%%%%%%%%%%%%%%%%%%%%%%%%%%%%%%%%%

We acknowledge significant support to NuGrid from STFC (through the University of Hull's Consolidated Grant ST/R000840/1), and access to {\sc viper}, the University of Hull High Performance Computing Facility. Thanks from JS go to the support provided by the University of Hull Internal Internship Programme. MP thanks the following for their support: the National Science Foundation (NSF, USA) under grant No. PHY-1430152 (JINA Center for the Evolution of the Elements), the "Lendulet-2014" Program of the Hungarian Academy of Sciences (Hungary), the ERC Consolidator Grant (Hungary) funding scheme (Project RADIOSTAR, G.A. n. 724560), the ChETEC COST Action (CA16117), supported by the European Cooperation in Science and Technology, the US IReNA Accelnet network (Grant No. OISE-1927130), and the European Union’s Horizon 2020 research and innovation programme (ChETEC-INFRA -- Project no. 101008324).

\section{Data Availability} %%%%%%%%%%%%%%%%%%%%%%%%%%%%%%%%%%%%%%%%%

The stellar data %underlying 
used for this article will be shared on reasonable request to the corresponding author. Stellar data files %used for this work 
are in hdf5 format, and tools to load and use the data are available on the public NuGrid github repository (https://github.com/NuGrid/NuGridPy). Presolar grains data are publicly available at the open source Presolar Grain Database (https://presolar.physics.wustl.edu/presolar-grain-database/).



\bibliographystyle{mnras}
\bibliography{main} % if your bibtex file is called example.bib

%%%%%%%%%%%%%%%%%%%%%%%%%%%%%%%%%%%%%%%%%%%%%%%%%%

%%%%%%%%%%%%%%%%% APPENDICES %%%%%%%%%%%%%%%%%%%%%

\appendix

\section{Additional Figures: $\delta$($^{47}$Ti/$^{48}$Ti) and $\delta$($^{50}$Ti/$^{48}$Ti)}

\begin{figure*} %% Ti47/48vs46/48
    \centering
    \includegraphics[width=1.0\textwidth]{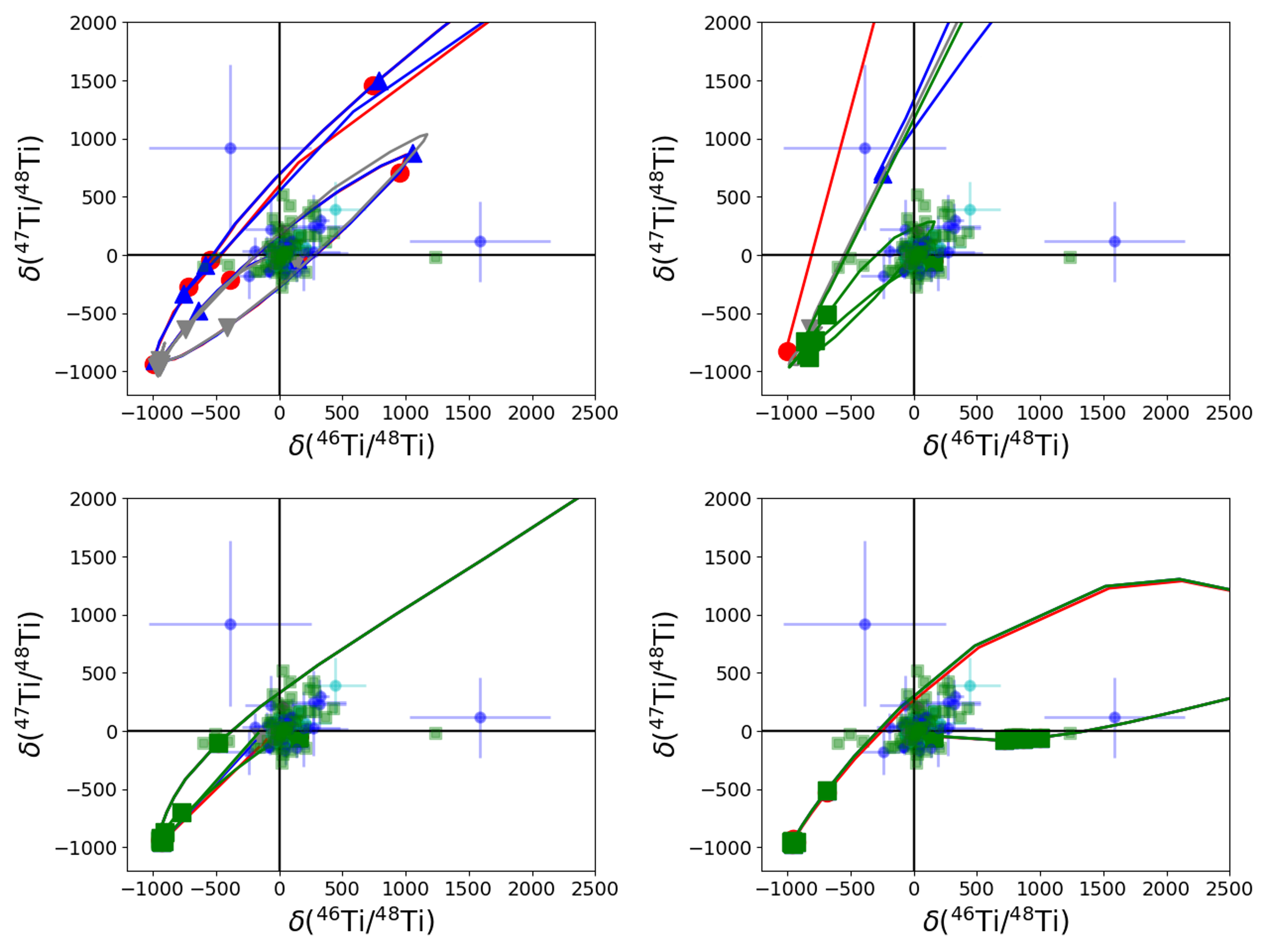}
    \caption{$^{47}$Ti/$^{48}$Ti and $^{46}$Ti/$^{48}$Ti isotopic ratios in $\delta$-notation of individual SiC grains (sub-types AB, nova, X, and C) and LD grains compared with predictions of models across the He/C zone. The C/Si zone of the 15d and 25T sets continue off the plots to the upper right of the region shown. For abundance ratios measured in grains and abundance profiles from stellar models symbols are used as in Figure \ref{fig:CvsN} in the four panels. Strong abundance variations are seen in the C/Si zone and the He/C zone, ejecting material with depletion of $^{47}$Ti with respect to $^{48}$Ti in some regions of the C/Si zone, or with strong production of $^{47}$Ti in the He/C zone.}
    \label{fig:Ti47}
\end{figure*}

\begin{figure*} %% Ti50/48 vs 46/48
    \centering
    \includegraphics[width=1.0\textwidth]{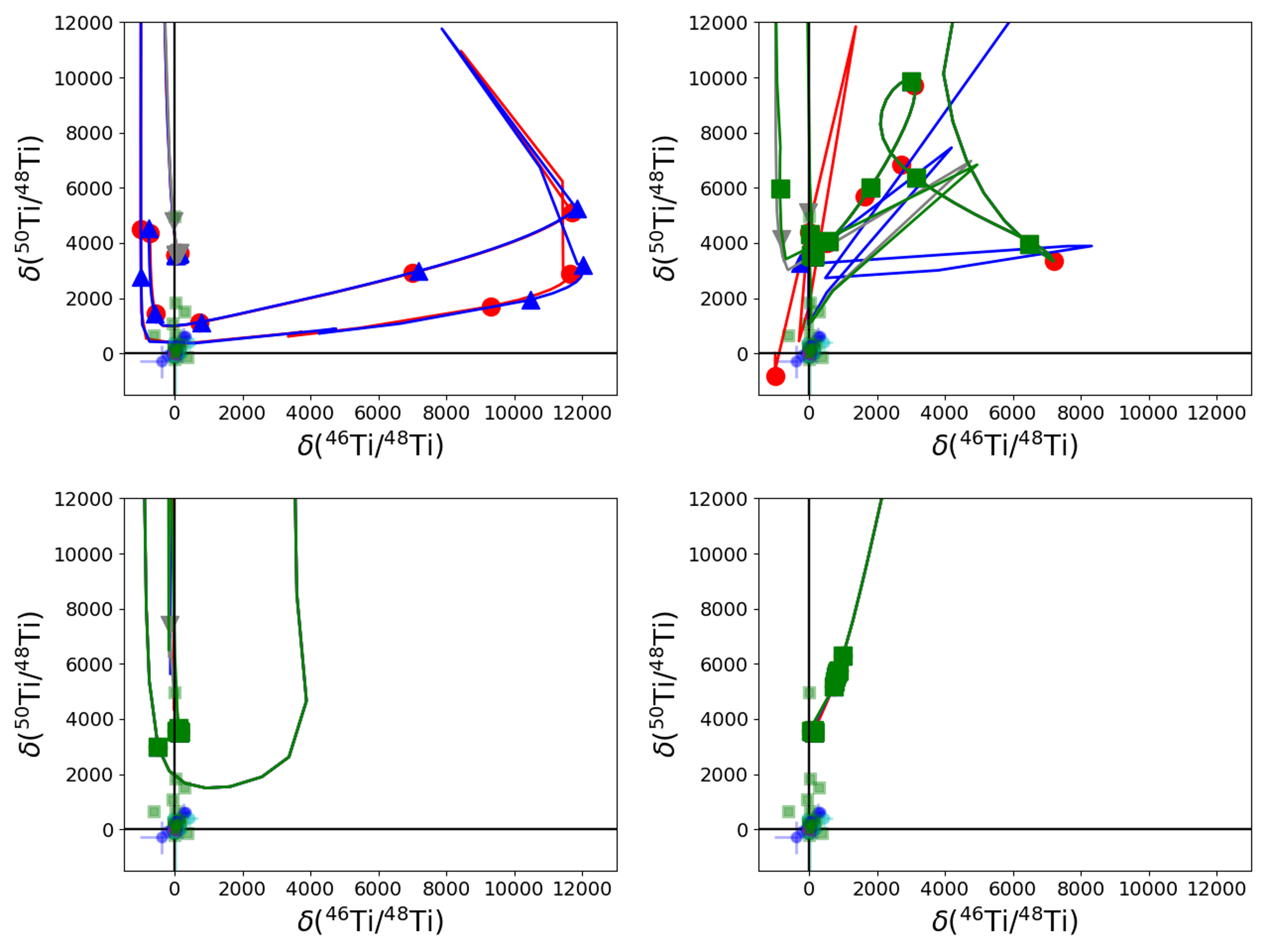}
    \caption{$^{50}$Ti/$^{48}$Ti and $^{46}$Ti/$^{48}$Ti isotopic ratios in $\delta$-notation shown for individual SiC grains (sub-types AB, nova, X, and C) and LD grains compared with predictions of models. For abundance ratios measured in grains and abundance profiles from stellar models symbols are used as in Figure \ref{fig:CvsN} in the four panels. As we have seen for $^{49}$Ti in Figure \ref{fig:Ti49}, high $^{50}$Ti/$^{48}$Ti is made by three main channels: in the C/Si zone, $^{50}$Ti stable isobars $^{50}$V and $^{50}$Cr are mostly made (but they cannot be distinguished from $^{50}$Ti in presolar grain measurements); in the O/Nova zone, as $^{50}$V from proton capture on $^{49}$Ti or as $^{50}$Cr from the proton-capture chain $^{48}$Ti(p,$\gamma$)$^{49}$V(p,$\gamma$)$^{50}$Cr; by neutron captures in the He/C zone, and $^{50}$Ti is mostly ejected as itself. In the figure no fractionation is applied when adding the $^{50}$Ti, $^{50}$V and $^{50}$Cr signatures. The highly abundant initial $^{50}$Cr is boosting the $\delta$($^{50}$Ti/$^{48}$Ti) to almost 4000 per mil even in the more external layers of the He/C zone, shown in the figure with normal $^{46}$Ti/$^{48}$Ti ratios.} 
    \label{fig:Ti50}
\end{figure*}

% If you want to present additional material which would interrupt the flow of the main paper,
% it can be placed in an Appendix which appears after the list of references.

%%%%%%%%%%%%%%%%%%%%%%%%%%%%%%%%%%%%%%%%%%%%%%%%%%

% Don't change these lines
\bsp	% typesetting comment
\label{lastpage}
\end{document}